\documentclass[12pt,a4paper]{article}

\usepackage{amsmath,amssymb}
\usepackage[nosort]{cite}
\usepackage[hyperref,bulletsep]{collect}
\def\gfxon{\usepackage[final]{graphicx}}

\gfxon

\makeatletter \@addtoreset{equation}{section} \makeatother

\makeatletter
\let\old@makecaption=\@makecaption
\def\@makecaption{\small\old@makecaption}
\makeatother

\makeatletter
\let\old@startsection=\@startsection
\renewcommand{\@startsection}[6]{\old@startsection{#1}{#2}{#3}{#4}{#5}{#6\mathversion{bold}}}
\makeatother

\let\oldPhi=\Phi
\let\oldPsi=\Psi
\let\oldGamma=\Gamma
\let\oldLambda=\Lambda
\let\oldDelta=\Delta
\let\oldSigma=\Sigma
\let\oldTheta=\Theta
\let\oldPi=\Pi
\renewcommand{\Phi}{\mathnormal{\oldPhi}}
\renewcommand{\Psi}{\mathnormal{\oldPsi}}
\renewcommand{\Gamma}{\mathnormal{\oldGamma}}
\renewcommand{\Sigma}{\mathnormal{\oldSigma}}
\renewcommand{\Delta}{\mathnormal{\oldDelta}}
\renewcommand{\Theta}{\mathnormal{\oldTheta}}
\renewcommand{\Lambda}{\mathnormal{\oldLambda}}
\renewcommand{\Pi}{\mathnormal{\oldPi}}

\newcommand{\superN}{\mathcal{N}}

\newcommand{\Op}{\mathcal{O}}
\newcommand{\Tr}{\mathop{\mathrm{Tr}}}

\newcommand{\order}[1]{\mathcal{O}(#1)}

\ifx\genfrac\sdflkaj
\newcommand{\atopfrac}[2]{{{#1}\above0pt{#2}}}
\else
\newcommand{\atopfrac}[2]{\genfrac{}{}{0pt}{}{#1}{#2}}
\fi
\newcommand{\sfrac}[2]{{\textstyle\frac{#1}{#2}}}
\newcommand{\half}{\sfrac{1}{2}}

\newcommand{\quarter}{\sfrac{1}{4}}

\newcommand{\indup}[1]{_{\mathrm{#1}}}

\newcommand{\alg}[1]{\mathfrak{#1}}

\newcommand{\lrbrk}[1]{\left(#1\right)}
\newcommand{\bigbrk}[1]{\bigl(#1\bigr)}

\makeatletter
\newcommand{\ellSN}{\mathop{\operator@font sn}\nolimits}
\newcommand{\ellCN}{\mathop{\operator@font cn}\nolimits}
\newcommand{\ellDN}{\mathop{\operator@font dn}\nolimits}
\newcommand{\ellAM}{\mathop{\operator@font am}\nolimits}
\newcommand{\ellK}{\mathop{\smash{\operator@font K}\vphantom{a}}\nolimits}
\newcommand{\ellE}{\mathop{\smash{\operator@font E}\vphantom{a}}\nolimits}
\newcommand{\gammafn}{\mathop{\smash{\oldGamma}\vphantom{a}}\nolimits}
\newcommand{\hypergeofn}[2]{\mathop{_{#1}\smash{\operator@font F}_{#2}\vphantom{a}}\nolimits}
\newcommand{\besselJ}[1]{\mathop{\smash{J_{#1}}\vphantom{a_{#1}}}\nolimits}
\newcommand{\bernoulli}{\mathrm{B}}
\renewcommand{\digamma}{\mathop{\smash{\oldPsi}\vphantom{a}}\nolimits}
\renewcommand{\Re}{\mathop{\operator@font Re}\nolimits}
\renewcommand{\Im}{\mathop{\operator@font Im}\nolimits}
\makeatother

\newcommand{\nn}{\nonumber}
\newcommand{\nln}{\nonumber\\}
\newcommand{\nl}[1][0pt]{\nonumber\\[#1]&\hspace{-4\arraycolsep}&\mathord{}}
\newcommand{\nlnum}{\\&\hspace{-4\arraycolsep}&\mathord{}}
\newcommand{\earel}[1]{\mathrel{}&\hspace{-2\arraycolsep}#1\hspace{-2\arraycolsep}&\mathrel{}}
\newcommand{\eq}{\earel{=}}

\def\[{\begin{equation}}
\def\]{\end{equation}}
\def\<{\begin{eqnarray}}
\def\>{\end{eqnarray}}
\def\be{\begin{equation}}
\def\ee{\end{equation}}
\def\ba{\begin{eqnarray}}
\def\ea{\end{eqnarray}}

\makeatletter
\def\mr@ignsp#1 {\ifx\:#1\@empty\else #1\expandafter\mr@ignsp\fi}%
\newcommand{\multiref}[1]{\begingroup
\xdef\mr@no@sparg{\expandafter\mr@ignsp#1 \: }%
\def\mr@comma{}%
\@for\mr@refs:=\mr@no@sparg\do{\mr@comma\def\mr@comma{,}\ref{\mr@refs}}%
\endgroup}
\makeatother

\newcommand{\hypref}[2]{\ifx\href\asklfhas #2\else\href{#1}{#2}\fi}

\newcommand{\secref}[1]{Sec.~\multiref{#1}}

\newcommand{\appref}[1]{App.~\multiref{#1}}

\newcommand{\figref}[1]{Fig.~\multiref{#1}}
\renewcommand{\eqref}[1]{(\multiref{#1})}

\ifx\href\asklfhas\newcommand{\href}[2]{#2}\fi
\newcommand{\arxivno}[1]{\href{http://arxiv.org/abs/#1}{#1}}

\begin{document}
\thispagestyle{empty}
\begin{flushright}\footnotesize
\texttt{\arxivno{hep-th/0610251}}\\
\texttt{AEI-2006-079}\\
\texttt{ITP-UU-06/44}\\
\texttt{SPIN-06/34}%
\vspace{0.5cm}
\end{flushright}
\vspace{0.1cm}

\renewcommand{\thefootnote}{\fnsymbol{footnote}}
\setcounter{footnote}{0}

\begin{center}%
{\Large\textbf{\mathversion{bold}%
Transcendentality and Crossing%
}}\vspace{1cm}

\textsc{Niklas Beisert$^{a}$, Burkhard Eden$^{b}$ and Matthias Staudacher$^{a}$} \vspace{5mm}

\textit{$^{a}$ Max-Planck-Institut f\"ur Gravitationsphysik\\
Albert-Einstein-Institut\\
Am M\"uhlenberg 1, D-14476 Potsdam, Germany}\vspace{3mm}

\textit{$^{b}$ Institute for Theoretical Physics and Spinoza Institute\\
Utrecht University\\
Postbus 80.195, 3508 TD Utrecht, The Netherlands}\vspace{3mm}

\texttt{nbeisert@aei.mpg.de}\\
\texttt{B.Eden@phys.uu.nl}\\
\texttt{matthias@aei.mpg.de}
\par\vspace{1cm}

\vspace{1.5cm}

\textbf{Abstract}\vspace{5mm}

\begin{minipage}{13.7cm}
We discuss possible phase factors for the
S-matrix of planar $\superN=4$ gauge theory,
leading to modifications at four-loop order as compared
to an earlier proposal. While these result in a four-loop
breakdown of perturbative BMN-scaling,
Kotikov-Lipatov transcendentality in the universal
scaling function for large-spin twist operators may be preserved.
One particularly natural choice, unique up to one
constant, modifies the overall contribution of all terms
containing odd zeta functions in the earlier proposed
scaling function based  on a trivial phase. Excitingly, we present
evidence that this choice is non-perturbatively related to
a recently conjectured crossing-symmetric phase factor
for perturbative string theory on  $AdS_5\times S^5$ once
the constant is fixed to a particular value.
Our proposal, if true, might therefore resolve the long-standing
AdS/CFT discrepancies between gauge and string theory.
\end{minipage} 

\vspace*{\fill}

\end{center}

\newpage
\setcounter{page}{1}
\renewcommand{\thefootnote}{\arabic{footnote}}
\setcounter{footnote}{0}

\section{Introduction}
\label{sec:intro}

Supersymmetric $\superN=4$ gauge theory is a conformal
quantum field theory in four dimensions. As such, standard lore
says that it does not possess an S-matrix since there are no
asymptotic particles. Luckily, this naive ``no-go'' theorem is
wrong in two interesting ways. 

Firstly, it turns out that an asymptotic ``world-sheet''
S-matrix of this theory 
may be defined, in the planar limit, in an internal space \cite{Staudacher:2004tk}. This is possible since planar local
composite operators may be interpreted as a one-dimensional 
ring on which 8 elementary bosons and 8 elementary fermions
propagate \cite{Berenstein:2002jq}. Furthermore, this S-matrix
appears to be \emph{two-particle factorized}, since the particle
model is integrable, at one-loop and beyond
\cite{Minahan:2002ve, Beisert:2003yb,Beisert:2003tq}.
This allows, under some assumptions, to largely construct the 
full S-matrix \cite{Beisert:2005tm}, up to an \emph{a priori}
unknown phase factor, and find the spectrum
of the theory \cite{Beisert:2005fw}. This phase factor
is known to equal one up to three-loop order.

Secondly, it is very interesting to consider the set of all space-time 
on-shell $n$-gluon amplitudes of the $\superN=4$ gauge theory.
While these are clearly not scattering amplitudes of 
asymptotic particles, they contain much physical information,
and we can simply declare them to serve as a surrogate
``space-time'' S-matrix. Fascinatingly, much evidence for an
iterative ``solvable'' structure in these amplitudes was 
discovered recently \cite{Anastasiou:2003kj,Bern:2005iz}.

The factorized world-sheet S-matrix and the iterative
space-time S-matrix approaches are not uncorrelated.
The conjectured ansatz for the all-loop gluon amplitudes 
contains a to-be-determined function $f(g)$, where
\[
g=\frac{\sqrt{\lambda}}{4\pi}\,
\]
is the gauge coupling constant. However, this function
$f(g)$ may be extracted from the logarithmically divergent 
anomalous dimension of leading-twist operators in the gauge theory
\cite{Sterman:2002qn}. As such, it may in principle be
derived from the diagonalization of the world-sheet S-matrix.
In this work we will call $f(g)$ the universal scaling function
of $\superN=4$ gauge theory; it is also known
as the cusp or soft anomalous dimension.

In practice, the comparison  has so far been done reliably 
to three-loop order only. The space-time based approach
was completed in \cite{Bern:2005iz}, and the world-sheet
method was exploited in \cite{Staudacher:2004tk}.
The results agreed. The latter method was also
successfully checked against rigorously known
anomalous dimensions of twist operators at
two loops, and brilliantly conjectured ones at three loops
\cite{Kotikov:2004er}. The conjecture was based on
an intriguing \emph{transcendentality principle}, which allowed  
to extract the answer from a hard QCD computation
\cite{Moch:2004pa}.

Clearly, one would like to go to higher loops, and ideally
compare the full universal scaling function. 
A four-loop computation by Bern, Czakon, Dixon, Kosower
and Smirnov using the iterative gluon amplitude approach
is under way \cite{Bern:2006aa,Kosower:2006aa}.%
\footnote{While finalizing our manuscript
we were informed that this computation
\protect\cite{Bern:2006ew} has been completed.}
Furthermore, in \cite{Eden:2006rx} it was shown how to extract
the universal scaling function from the Bethe equations
which result from the diagonalized S-matrix. Assuming a 
trivial (i.e.~equal to unity) phase factor, an all-loop 
equation determining the function was proposed.
While the exact solution of the equation remains unknown,
it allows to generate perturbative results to any desired order
with ease.  In addition, the solution preserves the Kotikov-Lipatov 
transcendentality principle \cite{Kotikov:2002ab}
to arbitrary loop order.

However, one troubling issue is whether the phase factor
is really unity to all orders in perturbation theory.
This leads to various vexing discrepancies 
\cite{Callan:2003xr,Serban:2004jf}
with string theory results 
in the context of the AdS/CFT correspondence
\cite{Maldacena:1998re,Gubser:1998bc,Witten:1998qj}. 
In fact, the S-matrix 
\cite{Beisert:2005tm,Hofman:2006xt,Arutyunov:2006ak}
of string theory on $AdS_5 \times S^5$
definitely requires a phase factor
\cite{Arutyunov:2004vx}
as can be seen by comparison to 
an integral equation describing classical solutions \cite{Kazakov:2004qf}.
Furthermore, it was argued that the S-matrix should
possess a \emph{crossing symmetry} \cite{Janik:2006dc},
which then necessarily calls for such a phase. 
In \cite{Arutyunov:2006iu} it was shown that the
known classical and one-loop part of the phase factor 
on the string side satisfies crossing symmetry, and in 
\cite{Beisert:2006ib} a crossing-invariant
all-order string phase factor was proposed.

In the weakly coupled gauge theory, the dressing phase
could appear, at the earliest, at four-loop order
\cite{Beisert:2005wv}. 
The leading (four-loop) effect on the scaling function
was already investigated in \cite{Eden:2006rx}. Here we
will extend this analysis to all-loop order, and investigate
the effects of the dressing phase on the universal scaling
function.

\section{Integral Equations for the Scaling Function}
\label{sec:dressedES}

Let us start in this section by considering 
the effect of the dressing phase on the universal scaling function. 
We first briefly recall the assumptions and main steps leading to an 
integral equation for 
the scaling function of leading twist operators of 
$\superN=4$ gauge theory in the large spin limit $S \rightarrow \infty$. 
Up to three loops, it is based on the asymptotic Bethe ansatz
\[
\label{allloopbethebs}
\left(\frac{x^+_k}{x^-_k}\right)^L=
\prod_{\textstyle\atopfrac{j=1}{j\neq k}}^S
\frac{x_k^--x_j^+}{x_k^+-x_j^-}\,
\frac{1-g^2/x_k^+x_j^-}{1-g^2/x_k^-x_j^+}\, .
\]
This ansatz was shown to properly work in the 
$\alg{sl}(2)$ sector relevant to twist operators up to three-loop
order in \cite{Staudacher:2004tk},
and further tests were performed in 
\cite{Eden:2005bt,Zwiebel:2005er}.
It was constructed to 
possibly describe gauge theory at four loops and
beyond in \cite{Beisert:2005fw}, 
in analogy with an earlier proposal for the $\alg{su}(2)$ sector \cite{Beisert:2004hm}.

The large-spin computation is similar to a thermodynamic limit
for a spin chain Bethe ansatz, as the number $S$ of Bethe roots
gets very large, and their distribution may be described
by a smooth density. Note, however, that the actual length $L$ of
this spin chain should remain quite short. Ideally it 
should be $L=2$, but this is dangerous since the Bethe
ansatz is a priori only asymptotic 
\cite{Serban:2004jf,Beisert:2004hm,Staudacher:2004tk,Beisert:2005fw}.
A work-around was devised in \cite{Eden:2006rx},
where it was argued that the scaling of the anomalous
dimension of the lowest state of operators of finite twist-$L$ is
a universal function of the coupling constant $g$.
This suggests that it should coincide with the scaling
function of twist-two operators.%
\footnote{At four loops the
asymptotic Bethe ans\"atze are only
fully reliable at twist three, or higher, and we have to
assume the four-loop universality of the scaling function for
the lowest state of twist operators. This universality may only 
be rigorously proved for twist $\geq 2$ at one loop 
\protect\cite{Belitsky:2006en}, for twist $\geq 2$ at two and three
loops, and for twist $\geq 3$ at four loops \protect\cite{Eden:2006rx}.
It would clearly be most welcome to more directly confirm
universality from the field theory.}
For a similar discussion,
albeit at one loop, see \cite{Belitsky:2006en}.

Despite the similarity to a thermodynamic limit, the computation
is quite subtle due to the singular distribution of the
one-loop roots \cite{Korchemsky:1995be}. 
The problem was solved in \cite{Eden:2006rx} by splitting
off the one-loop piece, and subsequently
deriving an integral equations for the non-singular 
fluctuations around it. It was found that the
density of fluctuations $\hat \sigma(t)$ is determined by the
solution of a non-singular integral equation of the general form
\[\label{fredholm}
\hat \sigma(t) \, = \, \frac{t}{e^{\, t } - 1} \; \Bigl[ \;
\, \hat K(2\, g \, t,0) -
4\, g^2 \int_0^\infty dt' \; \hat K(2\, g \, t, 2\, g \, t') \;
\hat \sigma(t') \,
\Bigr]\, .
\]
The universal scaling function $f(g)$ is then given by
\[\label{scalingfunction}
f(g) \, = \, 8 \, g^2 \, - \, 64 \, g^4 \, \int_0^\infty \,
dt\, \hat \sigma(t)\,
\frac{\besselJ{1}(2\, g\, t)}{2\, g \, t} \, .
\]
Here and in the following $\besselJ{r}(t)$ denotes a standard Bessel function.
Note that this expression can be reduced 
provided that the kernel satisfies the property 
(it will always be true in this article)
\[\label{dualpotential}
\hat K(0,t)=\frac{\besselJ{1}(t)}{t}\, .
\]
The scaling function then simply takes the value of the fluctuation density at $t=0$
\cite{Lipatov:2006aa}
\[\label{zero}
f(g)=16\,g^2\,\hat\sigma(0)\, .
\]

For the above, particular Bethe ansatz \eqref{allloopbethebs} 
the kernel $\hat K(t,t')=\hat K\indup{m}(t,t')$ is given by
(m stands for \emph{main} scattering) 
\[\label{mainkernel}
\hat K\indup{m}(t, t') =  
\frac{\besselJ{1}(t)  \besselJ{0}(t')  - \besselJ{0}(t)  \besselJ{1}(t')}{t - t'}\, .
\]
The inhomogeneous piece of the integral equation
\eqref{fredholm} with this kernel thus reads 
\[\label{potential}
\hat K\indup{m}(t,0)=\frac{\besselJ{1}(t)}{t}\, .
\]
For this specific kernel \eqref{mainkernel} derived from 
\eqref{allloopbethebs} the weak-coupling expansion of
the scaling function reads
\<\label{oldscalingshort}
   f\indup{m}(g)\eq 
8 g^2  
  -\frac{8}{3}\, \pi^2 g^4 
  +\frac{88}{45}\, \pi^4  g^6 
  - 16\lrbrk{\frac{73}{630}\,\pi^6 - 4\, \zeta(3)^2 }g^8
\nl 
  + 32 \lrbrk{\frac{887}{14175}\, \pi^8
              -\frac{4}{3}\, \pi^2 \zeta(3)^2
              -40\,\zeta(3)\, \zeta(5) } g^{10} 
\nl
- 64\,\biggl(
    \frac{136883}{3742200}\, \pi^{10}
  -\frac{8}{15}\,\pi^4\zeta(3)^2
  -\frac{40}{3}\,\pi^2\zeta(3)\,\zeta(5) 
\nl\qquad\quad
  -210\, \zeta(3)\, \zeta(7) 
  -102\, \zeta(5)^2 \biggr)\, g^{12} 
\nl
+128\,\biggl(
    \frac{7680089}{340540200}\, \pi^{12}
-\frac{47}{189}\, \pi^6\zeta(3)^2 
-\frac{82}{15}\, \pi^4 \zeta(3)\, \zeta(5) 
-70 \pi^2 \zeta(3)\,\zeta(7) 
\nl\qquad\qquad
-34 \pi^2 \zeta(5)^2 
-1176\, \zeta(3)\,\zeta(9) 
-1092\, \zeta(5)\,\zeta(7)
+4\, \zeta(3)^4 \biggr)\, g^{14}
\nl
  \mp  \ldots\, .
\>
Here we can observe the Kotikov-Lipatov transcendentality principle \cite{Kotikov:2002ab}: 
We attribute a degree of transcendentality $k$ to the constants $\pi^k$ as well as to $\zeta(k)$.

\emph{The $\ell$-loop contribution to the universal scaling function $f(g)$ 
for $\superN=4$ gauge theory
has a uniform degree of transcendentality 
$2\ell-2$.}

The function $f\indup{m}(g)$ derived from the integral equation \eqref{fredholm}
obeys this rule \cite{Eden:2006rx}. A rigorous proof to all
orders in perturbation theory was given in
\cite{Lipatov:2006aa}.

\medskip

Let us next generalize the universal scaling function to the case of 
an arbitrary weak-coupling dressing phase.
The only restriction is that the phase
should merely modify the gauge theory Bethe ansatz at four loops or beyond,
as the Bethe ansatz \eqref{allloopbethebs} is firmly established
up to three-loop order \cite{Beisert:2003ys,Eden:2004ua,Beisert:2005wv}. 
The corrections of the higher-loop 
Bethe equations which are consistent with current knowledge are
\[\label{allloopbethemod}
\left(\frac{x^+_k}{x^-_k}\right)^L=
\prod_{\textstyle\atopfrac{j=1}{j\neq k}}^S
\frac{x_k^--x_j^+}{x_k^+-x_j^-}\,
\frac{1-g^2/x_k^+x_j^-}{1-g^2/x_k^-x_j^+}\,
\exp\bigbrk{2i\theta(u_k,u_j)},
\]
where the dressing phase $\theta$
is conjectured to be of the general form
\cite{Arutyunov:2004vx,Staudacher:2005aa,Beisert:2005wv}
\[\label{dressing}
\theta(u_k,u_j)=
\sum_{r=2}^\infty
\sum_{\nu=0}^\infty
\beta_{r,r+1+2\nu}(g) 
\bigbrk{q_r(u_k)\, q_{r+1+2\nu}(u_j)-q_r(u_j)\, q_{r+1+2\nu}(u_k) }\, ,
\]
The $q_r(u)$ are the eigenvalues of the conserved magnon charges, 
see \cite{Beisert:2004hm}.
The coefficient functions $\beta_{r,s}(g)$ expand
in powers of the coupling constant
\[\label{betaexpand}
\beta_{r,r+1+2\nu}(g)=
\sum_{\mu=\nu}^\infty
g^{2 r+2\nu+2\mu}\beta_{r,r+1+2\nu}^{(r+\nu+\mu)}
\]
with $\beta_{r,s}^{(\ell)}$ some numerical constants
in the same notation as in \cite{Beisert:2005wv}.
We should eliminate from the start the only coefficient $\beta_{2,3}^{(2)}=0$
contributing at three loops.
Note that in \cite{Eden:2006rx} only the leading four-loop correction 
\[\label{sigma4}
\exp\bigbrk{2i\theta(u_k,u_j)}=
\exp\bigbrk{i\,8\beta\,g^6\left(q_2(u_k)\, q_3(u_j)-q_3(u_k)\, q_2(u_j) \right)
+ \ldots}
\]
was treated. Thus the first constant in \eqref{dressing} is 
$\beta_{2,3}^{(3)}=4\beta$. 

The modified Bethe ansatz \eqref{allloopbethemod}
with \eqref{dressing} may be treated in much the
same fashion as the simpler ansatz \eqref{allloopbethebs},
see \cite{Eden:2006rx} for the details. In particular,
the singular one-loop distribution may be split off in
the same way. 
In fact, for an arbitrary dressing phase
the general form \eqref{fredholm} of the integral equation 
as well as the properties \eqref{zero,dualpotential}
remain valid. What changes is the kernel \eqref{mainkernel},
which generalizes to 
\[\label{modkernel}
\hat K(t,t')=
\hat K\indup{m}(t,t')
+
\hat K\indup{d}(t,t')
\]
with the dressing kernel
\<\label{dresskernel}
\hat K\indup{d}(t,t')
\eq
\frac{4}{t\, t'}
\sum_{\rho=1}^\infty\sum_{\nu=0}^\infty\sum_{\mu=\nu}^\infty g^{2\mu+1}(-1)^\nu 
\Big(
\beta_{2\rho,2\rho+1+2\nu}^{(2\rho+\nu+\mu)}\besselJ{2\rho+2\nu}(t)\besselJ{2\rho-1}(t')
\nlnum\nonumber
\qquad\qquad\qquad\qquad\qquad\qquad\qquad
+\beta_{2\rho+1,2\rho+2\nu+2}^{(2\rho+1+\nu+\mu)}\besselJ{2\rho}(t)\besselJ{2\rho+1+2\nu}(t')
\Big)\, ,
\>
and the inhomogeneous term of the integral equation \eqref{fredholm} receives
additional contributions 
\[
\hat K\indup{d}(t,0)  =  \frac{\besselJ{1}(t)}{t}
+\frac{2}{t} 
\sum_{\nu=0}^\infty 
\sum_{\mu=\nu}^\infty
g^{2\mu+1} (-1)^\nu \,
\beta_{2,3+2\nu}^{(2+\nu+\mu)}\besselJ{2+2\nu}(t)\, .
\]
Notice that as promised \eqref{dualpotential} still holds for 
\eqref{modkernel,dresskernel} and arbitrary constants 
$\beta_{r,s}^{(\ell)}$, and that therefore also \eqref{zero} remains valid.

Equation \eqref{fredholm} is still just as suitable for a perturbative 
small-$g$ expansion if we use the modified kernel \eqref{modkernel,dresskernel} 
instead of \eqref{mainkernel}. It was already stated in \cite{Eden:2006rx} that 
a four-loop dressing phase \eqref{sigma4} leads to the following 
$\Op(g^8)$ term in the scaling function $f(g)$:
\[f(g)=\ldots
- 16\left( \frac{73}{630}\,\pi^6-4\,\zeta(3)^2 + 8 \, \beta \, \zeta(3)
\right)g^8 +\ldots\, .
\]
This violates transcendentality for generic $\beta$. However, if $\beta$ 
is a rational number times $\zeta(3)$ (or $\pi^3$) transcendentality is
\emph{preserved}. It is straightforward to extend this analysis to even 
higher loop order:

 \emph{Interestingly, a study of the effect of the constants 
$\beta_{r,s}^{(\ell)}$ in \eqref{dressing} reveals that for arbitrary loop order 
transcendentality is preserved if and only if the degree of transcendentality of
$\beta_{r,r+1+2\nu}^{(r+\nu+\mu)}$ is $2\mu+1$, independently of $r$ and $\nu$.
In other words} 
\[\label{eq:transbeta}
\beta_{r,s}^{(\ell)}\mbox{\emph{ should have degree of transcendentality }} 2\ell+2-r-s.
\medskip
\]

A particularly curious case is
$\beta=\frac{1}{2}\,\zeta(3)$, which cancels the term containing $\zeta(3)^2$, 
and thus leads to the much simpler four-loop answer $-\frac{73}{630}\,\pi^6\,16\,g^8$.
Beyond four loops, one finds that one can always choose the $\beta_{r,s}^{(\ell)}$
in many different ways such that \emph{all} terms containing zeta functions 
of \emph{odd} argument are canceled from the expansion \eqref{oldscalingshort}. 
What is truly remarkable, however, is that the constants $\beta_{r,s}^{(\ell)}$
are uniquely determined if we impose a further restriction on them 
conjectured to hold for arbitrary long-range spin chains
compatible with gauge theory%
\footnote{This study applied to the $\alg{su}(2)$ sector of the gauge theory, 
but the phase is universal for all sectors 
\protect\cite{Staudacher:2004tk,Beisert:2005fw,Beisert:2005tm}
and thus includes $\alg{sl}(2)$.} 
in \cite{Beisert:2005wv}, cf.~eq.~(4.2) in that paper,
\[\label{eq:feynmanconstraint}
\beta_{r,r+1+2\nu}^{(r+\nu+\mu)}=0 
 \qquad \mbox{for} \qquad 
\mu<r+\nu-1\, ,
\]
or, in a different notation, $\beta_{r,s}^{(\ell)}=0$ for $\ell<r+s-2$.
One finds to the first few orders%
\footnote{Some of the coefficients appear 
in the expansion of the scaling function at higher orders 
than naively expected from the expansion of the kernel 
\protect\eqref{modkernel,dresskernel}. 
They are nevertheless fixed.
E.g.~$\beta_{3,4}^{(5)}$ does not
appear at six loops as expected, but only at
seven loops in $f(g)$. But at this order it is fixed
to $12\,\zeta(5)$ if we demand all odd-zeta terms to cancel.}
\[
\label{cancelzeta}
\begin{array}[b]{rclrclrcl}
\beta_{2,3}^{(3)}\earel{\to}+2\,\zeta(3), \\[3pt]
\beta_{2,3}^{(4)}\earel{\to}-20\,\zeta(5), \\[3pt] 
\beta_{2,3}^{(5)}\earel{\to}+210\,\zeta(7),
&\beta_{3,4}^{(5)}\earel{\to}+12\,\zeta(5),
&\beta_{2,5}^{(5)}\earel{\to}-4\,\zeta(5) ,\\[3pt]
\beta_{2,3}^{(6)}\earel{\to}-2352\,\zeta(9),
&\beta_{3,4}^{(6)}\earel{\to}-210\,\zeta(7),
&\beta_{2,5}^{(6)}\earel{\to}+84\,\zeta(7).
\end{array}
\]
The expansion of the scaling function 
significantly simplifies as compared to \eqref{oldscalingshort}:
\<\label{scalingsym}
f_0(g) \eq 
 8 \, g^2 
   -\frac{8}{3} \, \pi ^2 \, g^4 
   + \frac{88}{45} \, \pi ^4 \, g^6 
   - 16\,\frac{73}{630}  \pi^6 \, g^8 
   + 32\,\frac{887}{14175} \, \pi^8 g^{10}
\nl
   - 64\,\frac{136883}{3742200} \, \pi ^{10} g^{12}
   + 128\,\frac{7680089}{340540200}\, \pi^{12} g^{14}
   \mp \, \ldots \, .
\>
The even zeta terms, and thus the parts containing only even powers
of $\pi$, are unaffected. 

\medskip

However, \eqref{cancelzeta} is not the only curious choice for the
constants $\beta_{r,s}^{(\ell)}$ in the dressing phase
\eqref{dressing}. Another striking choice corresponds to
doubling the just discussed special constants, 
e.g.~to the first few order \eqref{cancelzeta} become:
\[
\label{flipsigns}
\begin{array}[b]{rclrclrcl}
\beta_{2,3}^{(3)}\eq+4\,\zeta(3), \\[3pt]
\beta_{2,3}^{(4)}\eq-40\,\zeta(5), \\[3pt] 
\beta_{2,3}^{(5)}\eq+420\,\zeta(7),
&\beta_{3,4}^{(5)}\eq+24\,\zeta(5),
&\beta_{2,5}^{(5)}\eq-8\,\zeta(5) ,\\[3pt]
\beta_{2,3}^{(6)}\eq-4704\,\zeta(9),
&\beta_{3,4}^{(6)}\eq-420\,\zeta(7),
&\beta_{2,5}^{(6)}\eq+168\,\zeta(7).
\end{array}
\]
Now the zeta functions of odd argument no longer cancel
out. Instead, one finds to e.g.~seven-loop order 
\<\label{scalingshort}
   f(g)\eq 
8 g^2  
  -\frac{8}{3}\, \pi^2 g^4 
  +\frac{88}{45}\, \pi^4  g^6 
  - 16\lrbrk{\frac{73}{630}\,\pi^6 + 4\, \zeta(3)^2 }g^8
\nl 
  + 32 \lrbrk{\frac{887}{14175}\, \pi^8
              +\frac{4}{3}\, \pi^2 \zeta(3)^2
              +40\,\zeta(3)\, \zeta(5) } g^{10} 
\nl
- 64\,\biggl(
    \frac{136883}{3742200}\, \pi^{10}
  +\frac{8}{15}\,\pi^4\zeta(3)^2
  +\frac{40}{3}\,\pi^2\zeta(3)\,\zeta(5) 
\nl\qquad\quad
  +210\, \zeta(3)\, \zeta(7) 
  +102\, \zeta(5)^2 \biggr)\, g^{12} 
\nl
+128\,\biggl(
    \frac{7680089}{340540200}\, \pi^{12}
+\frac{47}{189}\, \pi^6\zeta(3)^2 
+\frac{82}{15}\, \pi^4 \zeta(3)\, \zeta(5) 
+70 \pi^2 \zeta(3)\,\zeta(7) 
\nl\qquad\qquad
+34 \pi^2 \zeta(5)^2 
+1176\, \zeta(3)\,\zeta(9) 
+1092\, \zeta(5)\,\zeta(7)
+4\, \zeta(3)^4 \biggr)\, g^{14}
\nl
  \mp  \ldots\, .
\>
Remarkably, the alternating sum
\eqref{scalingshort} is identical
to \eqref{oldscalingshort} for the
case of a trivial dressing phase by multiplying all 
zeta functions with odd arguments by 
the imaginary unit $i$, i.e.~the replacement $\zeta(2n+1) \rightarrow i\,\zeta(2 n+1)$.
After this operation,
and in contradistinction the the earlier case as discussed in
\cite{Eden:2006rx}, now all relative signs of the terms in 
\eqref{scalingshort} are identical,
but the terms are otherwise unchanged!
A proof of this transformation will be given in 
\appref{app:flipzeta}.

To wrap up the above results, 
we would like to mention that the scaling functions $f\indup{m}(g)$, $f_0(g)$ 
and $f(g)$ are part of a one-parameter family $f_\kappa(g)$
interpolating between these three choices. The general function is obtained by
multiplying the constants in \eqref{cancelzeta}
by an overall factor of $(1+\kappa)$. 
The resulting universal scaling function $f_\kappa(g)$ 
is the same as the sign-synchronized scaling function $f(g)$ in \eqref{scalingshort}, 
but with the replacement $\zeta(2n+1) \rightarrow \sqrt{\kappa}\,\zeta(2 n+1)$.

\medskip

We see that there are very interesting and seemingly
natural ways 
to deform the scattering phase of \cite{Beisert:2004hm,Beisert:2005fw}
while preserving Kotikov-Lipatov transcendentality. We will now argue in
section \ref{sec:analytic} that the (conjectured) AdS/CFT
correspondence, together with (conjectured)
integrability and a (conjectured) crossing-symmetric
strong-coupling phase factor indeed appears to single out 
one of the above choices. 
We will continue in \secref{sec:kernels} with the investigation
of the kernels and scaling functions. 
There we will derive a closed form for the very same constants 
and a concise integral expression for the summed dressing kernel.

\section{An Analytic Continuation of Sorts}
\label{sec:analytic}

In the following section, we shall investigate the
constants $\beta^{(\ell)}_{r,s}$ starting from string theory.
For perturbative string theory it is useful to write the dressing phase 
$\theta_{k,j}$ in \eqref{allloopbethemod,dressing} as 
\[
\theta(u_k,u_j)=\sum_{r=2}^\infty\sum_{s=r+1}^\infty c_{r,s}(g)
\bigbrk{\tilde q_{r}(u_k)\,\tilde q_{s}(u_j)-\tilde q_{s}(u_k)\,\tilde q_{r}(u_j)}
\]
Here the excitation charges $\tilde q_r(u)$ are normalized as 
$\tilde q_r(u)=g^{r-1}q_r(u)$
differently from \eqref{dressing}.
Consequently, the coefficient functions $c_{r,s}$ and $\beta_{r,s}$ are related by 
\[
c_{r,s}(g)=g^{2-r-s}\beta_{r,s}(g).
\]
The strong-coupling expansion of $c_{r,s}$ within string theory 
is non-trivial \cite{Arutyunov:2004vx,Beisert:2005cw,Hernandez:2006tk,Freyhult:2006vr}
\[\label{eq:cexp.strong}
c_{r,s}(g)=\sum_{n=0}^\infty c_{r,s}^{(n)} g^{1-n}.
\]
A proposal for the all-order strong-coupling expansion 
based on available data
\cite{Arutyunov:2004vx,Beisert:2005cw,Hernandez:2006tk,Freyhult:2006vr}
and crossing symmetry \cite{Janik:2006dc}
was made in Sec.~5.2 of \cite{Beisert:2006ib} 
\[\label{eq:coeffstrong}
c^{(n)}_{r,s}=\frac{\bigbrk{1-(-1)^{r+s}}\zeta(n)}{2(-2\pi)^n \gammafn(n-1)}\,(r-1)(s-1)\,
\frac{\gammafn[\half(s+r+n-3)]\gammafn[\half(s-r+n-1)]}
     {\gammafn[\half(s+r-n+1)]\gammafn[\half(s-r-n+3)]}\,.
\]
This expression is formally $0/0$ when setting $n=0,1$ 
so we have to use a proper regularization \cite{Beisert:2006ib}:
In these cases the correct expression is known from comparison to 
classical and one-loop string theory data \cite{Arutyunov:2004vx,Beisert:2005cw,Hernandez:2006tk,Freyhult:2006vr}
\[\label{eq:coeff01}
c^{(0)}_{r,s}=\delta_{r+1,s}
\,,\qquad
c^{(1)}_{r,s}=-\frac{\bigbrk{1-(-1)^{r+s}}}{\pi}\,
\frac{(r-1)(s-1)}{(s+r-2)(s-r)}\,.
\]
The expression for $c^{(1)}_{r,s}$ is easily recovered from
\eqref{eq:coeffstrong} through the limit
$\zeta(n)/\gammafn(n-1)\to 1$ when $n\to 1$.
For $n=0$ we first set $r,s$ to some integer values with $s>r$. 
If $s>r+1$ the only singular term is $1/\gammafn(n-1)$ 
which guarantees $c^{(0)}_{r,s}=0$.
This explains the $\delta_{r+1,s}$ term in \eqref{eq:coeff01}. 
For $s=r+1$, however, the term $\gammafn[\half(s-r+n-1)]$ diverges
at $n=0$. In this case we can rewrite the expression 
with arbitrary $n$ as
\[
c^{(n)}_{r,r+1}=\frac{2(n-1)\,\zeta(n)\,\gammafn(1+\half n)}{(-2\pi)^n \gammafn(n+1)\gammafn(2-\half n)}\,
\frac{r(r-1)\gammafn(r-1+\half n)}{\gammafn(r+1-\half n)}\,.
\]
Here we can easily set $n=0$ and obtain $c^{(0)}_{r,r+1}=1$
as required by \eqref{eq:coeff01}.

\medskip

In order to compare to gauge theory,
we have to find an expansion of $c_{r,s}(g)$ at weak
coupling according to \eqref{dressing}. 
The main obstruction in the investigation of the
function $c_{r,s}(g)$ is that the known expansion 
at strong coupling is merely asymptotic:
For fixed $r,s$ and even $n$ the sequence of coefficients
$c^{(n)}_{r,s}$ terminates at $n=s-r+1$ \cite{Beisert:2006ib}.
However, the odd-$n$ sequence does not terminate
and grows factorially as
\[
\frac{c^{(n+2)}_{r,s}}{c^{(n)}_{r,s}}=\frac{n^2}{64\pi^2}+\order{n}.
\]
Consequently, the series of $c^{(n)}_{r,s}$ is asymptotic 
and has zero radius of convergence around $g=\infty$.
It is Borel summable, but the expression
for the coefficients may be too complex to perform the sum in practice.

\medskip

In order to extrapolate to weak coupling, 
let us consider a simple model function first.
In \cite{Beisert:2006ib} some similarities
between the phase and the digamma function 
$\digamma(z)=\partial_z\log \gammafn(z)$ 
were observed.
This function has the following asymptotic expansion
for large and positive $z$
\[
\digamma(1+z)=\log z+\sum_{n=1}^\infty
\frac{c_n}{z^n}\,,\qquad
c_n=-\frac{\bernoulli_n}{n}=(-1)^{n}\zeta(1-n)\,,
\]
where $\bernoulli_n$ are the Bernoulli numbers
which can also be expressed through the zeta function as
$\bernoulli_n=(-1)^{n+1}n\zeta(1-n)$.
Conversely, the expansion around $z=0$ reads
\[
\digamma(1+z)=-\gamma\indup{E}
+\sum_{k=1}^\infty
\tilde c_k\,z^{k},\qquad
\tilde c_k=-(-1)^{k}\zeta(1+k)
\]
with $\gamma\indup{E}$ being Euler's constant.
The curious observation is that the expansion
coefficients for large $z$ and for small $z$ are almost the same.
They are related by
\[
c_n=-\tilde c_{-n}.
\]
In other words, we obtain the expansion coefficients
by \emph{analytic continuation}
of the index parameter $n$ to negative values.

\medskip

Could it be that a similar relation holds also 
for the expansion of the coefficients $c_{r,s}(g)$ 
at strong and weak coupling, respectively?
Admittedly, this is a very wild guess, but it turns out to be 
literally true. The expansion of $c_{r,s}(g)$ at weak coupling
reads
\[\label{eq:cexp.weak}
c_{r,s}(g)=-\sum_{n=1}^\infty c_{r,s}^{(-n)}g^{1+n}.
\]
In the following we shall not only provide indications
for the mathematical correctness of this statement,
but also argue that it leads to a consistent physical picture.

As a first test, let us compute some $c_{r,s}^{(n)}$
with negative value of $n$. 
Straight evaluation turns out to yield $0/0$ in most cases 
and the expressions need to be regularized.
We will therefore use the same procedure as in the cases $n=0,1$ and
set $r,s$ to their expected values first. Afterwards we
take the analytic continuation of $c_{r,s}^{(n)}$ for
the expected value of $n$. 
In this way we obtain, e.g. 
\[
c_{2,3}^{(-1)}=0\,,\qquad
c_{2,3}^{(-2)}=-4\,\zeta(3)\,.
\]
The absence of a $n=-1$ contribution translates via $\ell=(r+s-n-1)/2$
to the absence of a $\ell=5/2$ contribution, where $\ell$ represents
the number of loops in gauge theory. 
In general all contributions for odd negative $n$ should be zero 
to guarantee absence of contributions at fractional orders. 
In that case the expansion coefficients at weak coupling would read
\[\label{eq:WL}
\beta^{(\ell)}_{r,s}=-c_{r,s}^{(r+s-2\ell-1)}.
\]
The first non-trivial contribution appears at%
\footnote{A curious observation is that the leading contribution to
the function $\beta_{2,3}=\zeta(3)\,\lambda^3/1024\pi^6+\ldots$
resembles the value 
$3\zeta(3)\lambda^3/512\pi^4$ \protect\cite{Beisert:2002aa}
of a certain X-shaped diagram \protect\cite{Arutyunov:2001hs}
to a circular Maldacena-Wilson loop \protect\cite{Maldacena:1998im}
(the conjectured value $\lambda^3/4!^3$ in \protect\cite{Arutyunov:2001hs} is almost true).
If a contribution proportional to $\zeta(3)$ survives in the
sum over all diagrams at this order, it would imply
that the ladder approximation \cite{Erickson:2000af,Drukker:2000rr} is not complete.
This may seem unlikely, as this approximation 
yields a consistent strong-coupling result
\cite{Erickson:2000af,Drukker:2000rr}.
However, it is not yet excluded either that non-ladder contributions 
have a mild, but non-vanishing effect.}
\[\label{flipsign2}
\beta_{2,3}^{(3)}=4\,\zeta(3).
\]
This is encouraging for several reasons:
Firstly, it influences anomalous dimensions
starting from four loops. 
Currently, we know $\beta_{2,3}^{(\ell)}=0$ for $\ell=0,1,2$ only,
so this is perfectly consistent with available data from gauge theory.
Moreover, $\beta_{2,3}^{(3)}$ is the first allowed contribution 
according to \eqref{eq:feynmanconstraint}. 
In addition, the value $4\zeta(3)$ has transcendentality $3$ 
in the counting scheme of \cite{Kotikov:2002ab}. 
This is precisely the right transcendentality required for
a uniform transcendentality of the scaling function
\cite{Eden:2006rx}.
Finally, the value \eqref{flipsign2} agrees precisely with
the first in the series of sign reversing constants \eqref{flipsigns}!

Let us now rewrite the coefficients in a form more convenient for
negative $n$. We use the identities
\[
\zeta(1-z)=2(2\pi)^{-z}\cos(\half\pi z)\gammafn(z)\,\zeta(z)\quad
\mbox{and}\quad
\gammafn(1-z)=\frac{\pi}{\sin(\pi z)\gammafn(z)}
\]
and obtain for integer $r,s$ a new form for the coefficients
\[\label{result}
c^{(n)}_{r,s}=
\frac{\bigbrk{1-(-1)^{r+s}}\cos(\half\pi n)\,(-1)^{s-1-n}\,\zeta(1-n)\,\gammafn(2-n)\gammafn(1-n)\,(r-1)(s-1)}
{\gammafn[\half(5-n-r-s)]\gammafn[\half(3-n+r-s)]
 \gammafn[\half(3-n-r+s)]\gammafn[\half(1-n+r+s)]}\,.
\]
The factor $\cos(\half\pi n)$ makes it apparent that
$n$ must be even and that there are no fractional-loop contributions.
Furthermore the factor $1/\gammafn[\half(5-n-r-s)]$ imposes the 
lower bound $-n\geq r+s-3$ for the loop order
$\ell=(r+s-n-1)/2\geq r+s-2$
consistent with \eqref{eq:feynmanconstraint}.
As we discussed already in the previous section \ref{sec:dressedES},
\[
\beta^{(\ell)}_{r,s}\sim \zeta(2+2\ell-r-s)
\]
this is precisely what is required for a uniform transcendentality 
of the scaling function.

\medskip

\begin{figure}\centering
\includegraphics[width=7cm]{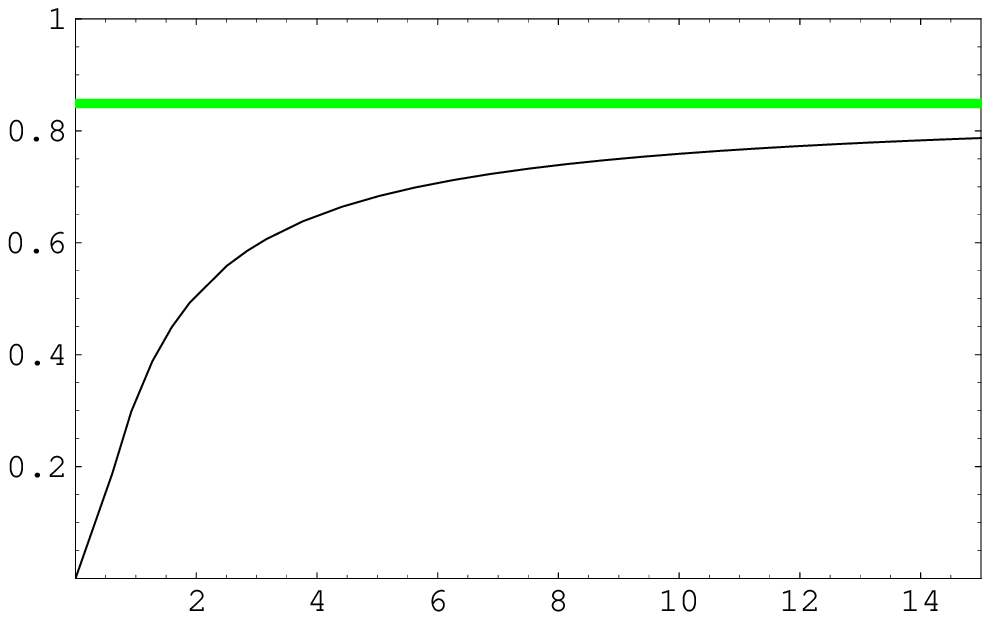}\qquad
\includegraphics[width=7cm]{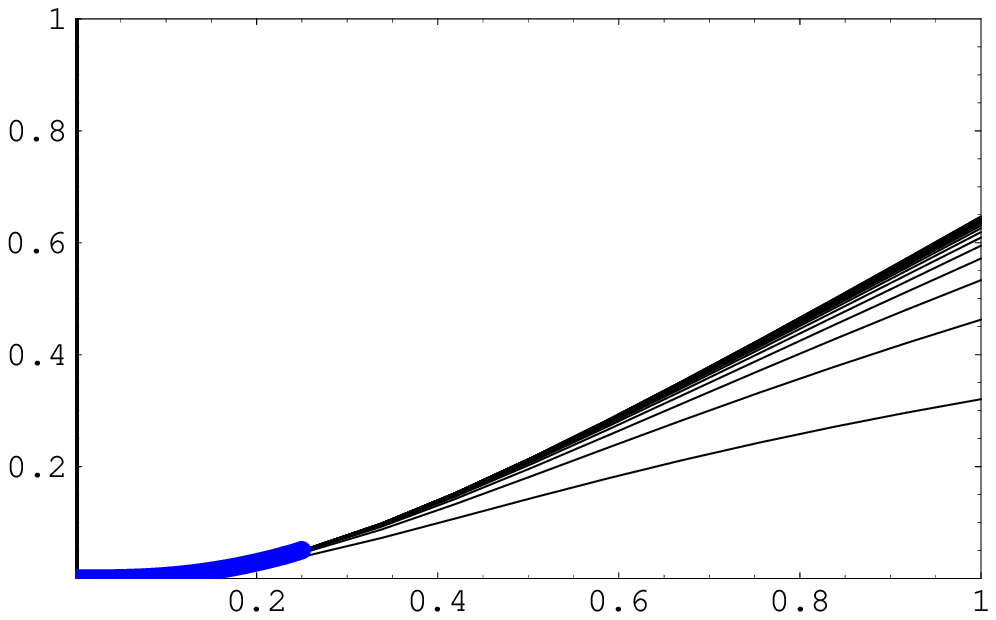}
\caption{Left: Plot of $h_{23}(z)$ with asymptotic
value $8/3\pi$. Right: Plot of the partial sums in $c_{23}(g)$
with weak coupling expansion up to
radius of convergence $g=\quarter$.}
\label{fig:weak}
\end{figure}

Being convinced of the physical plausibility of the proposal,
we shall now investigate the equivalence of 
the weak and strong-coupling expansions
for the choice $r=2,s=3$.
The proposed weak-coupling
expansion is 
\[
c_{2,3}(g)=-\sum_{n=1}^\infty c_{2,3}^{(-2n)}g^{1+2n}
\]
with 
\[
c^{(-2n)}_{2,3}=
\frac{4(-1)^n\,\zeta(2n+1)\,\gammafn(2n)\gammafn(2n-1)}
{\gammafn(n)\gammafn(n+1)\gammafn(n+2)\gammafn(n+3)}\,.
\]
We shall write $\zeta(1+2n)$ using its series representation
\[
c_{2,3}(g)=\sum_{n=1}^\infty \sum_{k=1}^\infty
\frac{4(-1)^{n+1}\,(2n-1)!\,(2n)!\,(g/k)^{1+2n}}
{(n-1)!\,n!\,(n+1)!\,(n+2)!}\,.
\]
The sum over $n$ can now be evaluated explicitly to 
\[\label{eq:ch23}
c_{2,3}(g)=\sum_{k=1}^\infty h_{2,3}(g/k)\,,\qquad
h_{2,3}(z)=-\frac{1}{z}
+\frac{1+16z^2}{6\pi z^3}\,\ellK(4iz)
-\frac{1-4z^2}{6\pi z^3}\,\ellE(4iz),
\]
where $\ellK$ and $\ellE$ are the elliptic integrals
of the first and second kind, respectively.%
\footnote{We use the convention that the  
modulus appears in a squared form as $(4iz)^2=m$.}
The function $h_{2,3}$, some partial sums 
as well as the series expansion of $c_{2,3}$ 
are displayed in \figref{fig:weak}.

\begin{figure}\centering
\includegraphics[width=7cm]{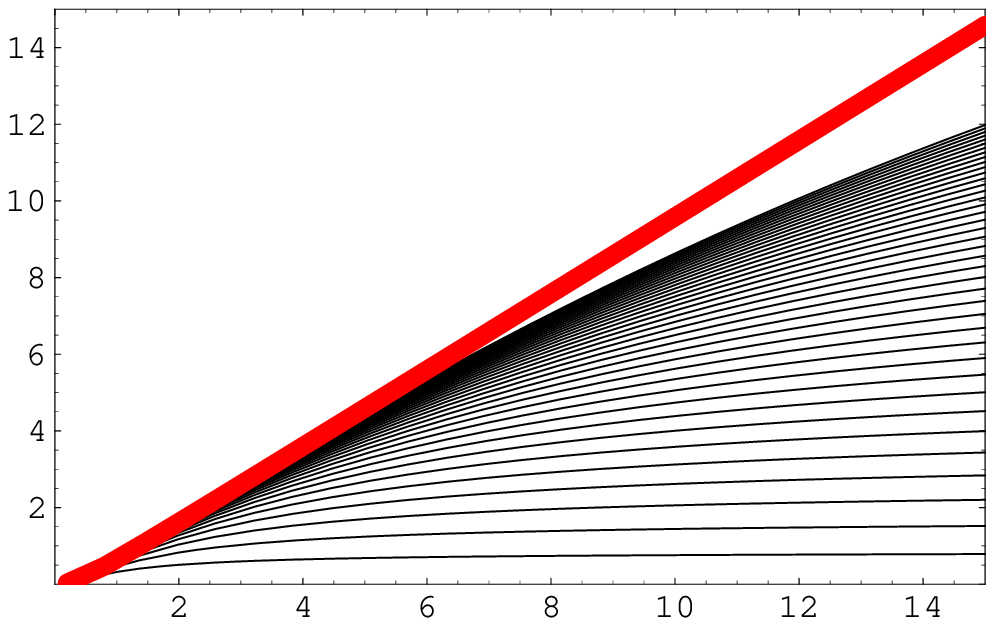}\qquad
\includegraphics[width=7cm]{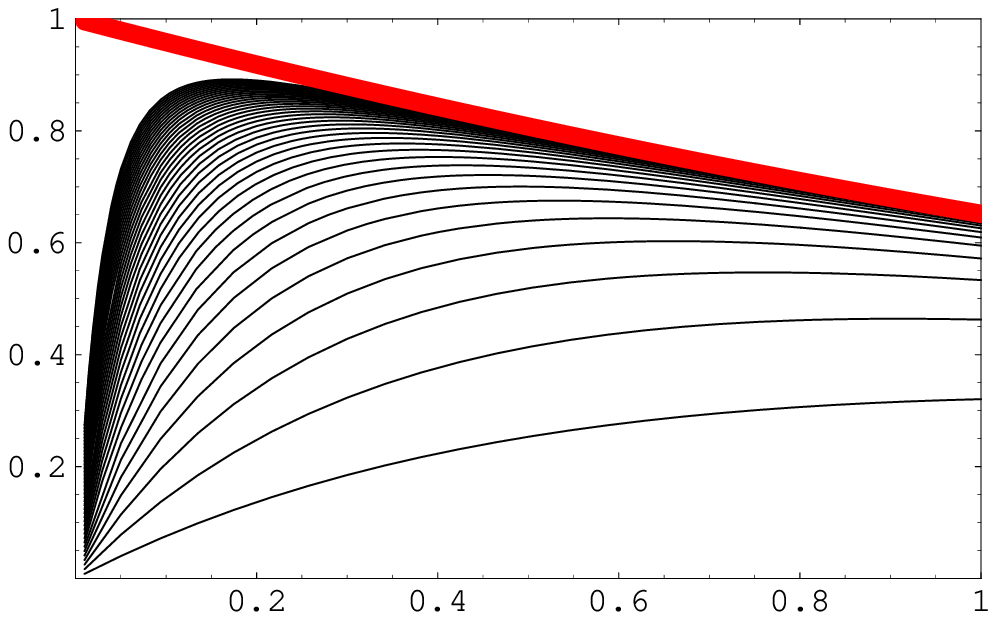}
\caption{Left: Plot of partial sums in $c_{2,3}(g)$ 
with asymptotic strong coupling expansion. 
Right: We have plotted $xc_{2,3}(1/x)$ instead to
improve the view of the strong-coupling behavior.}
\label{fig:strong}
\end{figure}

It is now possible to connect to strong coupling. 
The expansion of the function $h_{2,3}$ at large positive $z$ is
\<
h_{2,3}(z)=
\frac{8}{3\pi}
-\frac{1}{z}
+\frac{3\log(16z)}{4\pi z^2}
-\frac{5}{8\pi z^2}
+\frac{5\log(16z)}{512\pi z^4}
+\frac{1}{2048\pi z^4}
+\ldots\,.
\>
Zeta function regularization%
\footnote{Zeta function regularization
is closely related to Euler-MacLaurin summation.
However, the latter has to be adapted to deal correctly with terms $k^n\log(k)$ 
in the expansion of $h$, see \protect\cite{Celorrio:1998aa}.
We thank S.~Frolov for providing us with this reference.}
of the sum yields
\<\label{eq:c23zeta}
\sum_{k=1}^\infty h_{2,3}(g/k)\eq
g\int_0^\infty h_{2,3}(1/z)\,dz
+\frac{8\zeta(0)}{3\pi}
-\frac{\zeta(-1)}{g}
\nl
+\frac{3}{4\pi}\,\frac{\zeta'(-2)}{g^2}
-\frac{5}{8\pi}\,\frac{\zeta(-2)}{g^2}
+\frac{5}{512\pi}\,\frac{\zeta'(-4)}{g^4}
-\frac{1}{2048\pi}\,\frac{\zeta(-4)}{g^4}
+\ldots
\nln\eq
g-\frac{4}{3\pi}+\frac{1}{12g}
-
\frac{3\zeta(3)}{16\pi^3g^2}
+
\frac{15\zeta(5)}{2048\pi^5g^4}+\ldots
\nln\eq
\sum^\infty_{n=0}c^{(n)}_{2,3}g^{1-n}.
\>
Here the sum over $k^n$ is regularized to $\zeta(-n)$ 
while the sum over  
$k^n\log(k)$ yields $-\zeta'(-n)$.
We have verified the perfect agreement
of the expansion with the strong-coupling coefficients 
$c^{(n)}_{2,3}$ to order $g^{-100}$ 
leaving little room for error.
In \appref{sec:strong} we present a 
proof of this statement to all orders. 
In the proof we observe some exponential corrections 
of the type $\order{e^{-g}}$
to the series.
These corrections are not unexpected because the series is asymptotic.
A plot of the partial sums and asymptotic
strong-coupling expansion of $c_{2,3}$ is given in \figref{fig:strong}.

\medskip

We have also confirmed the same agreement
for other values of $(r,s)$:
We can always replace the $\zeta(1+2n)$ in $c_{r,s}$ by a new sum
and write $c_{r,s}$ in terms of a function $h_{r,s}(z)$ without 
odd-zeta terms%
\footnote{This form of the phase may be suggestive of the presence
of bound states \protect\cite{Dorey:2006dq,Chen:2006gq,Roiban:2006gs}
(or in general larger multiplets \protect\cite{Beisert:2006qh})
in intermediate channels during the scattering process.
For instance, the dispersion relation 
as well as other characteristic quantities
of a bound state of $k$ particles is typically
obtained by replacing $g\to g/k$ in the
expressions for a single particle.}
\[\label{csum}
c_{r,s}(g)=\sum_{k=1}^\infty h_{r,s}(g/k).
\]
The series in $g$ can now be written formally as
the hypergeometric function 
\[\label{hhyper}
h_{r,s}(z)=
\frac{2\cos[\half\pi(s-r-1)]\gammafn(r+s-2)}{\gammafn(r-1)\gammafn(s-1)}\,
z^{r+s-2}\,\hypergeofn{4}{3}(\vec{a};\vec{b};-16z^2)
\]
with arguments 
$\vec{a}=[
\half(r+s-2),
\half(r+s-1),
\half(r+s-1),
\half(r+s)]$ and $\vec{b}=[r,s,r+s-1]$.
This hypergeometric function can be evaluated to the general form 
using elliptic integrals
\[\label{hKEform}
h_{r,s}(z)=\frac{P_{r,s}(z^2)}{z^{r+s-2}}
+\frac{Q_{r,s}(z^2)}{\pi z^{r+s-2}}\,\ellK(4iz)
+\frac{R_{r,s}(z^2)}{\pi z^{r+s-2}}\,\ellE(4iz).
\]
Here $P,Q,R$ are some polynomials of degree $(r+s-3)/2$
with rational coefficients. The polynomials $P$ and $R$ have 
special factors for all value of $r,s$, namely $P(z^2)\sim z^{2(r-2)}$ and 
$Q(z^2)\sim (1+16z^2)$.
As in the above case of $(r,s)=(2,3)$, we use zeta function regularization 
to evaluate the sums \eqref{csum} at strong coupling.%
\footnote{The polynomial $P$ generates
even loop orders in string theory, 
while $Q,R$ generate the odd ones.
Its coefficients are thus directly related 
to the coefficients $c_{r,s}^{(2n)}$.}
We have further expanded some of the functions \eqref{hhyper}
at strong coupling and they fully agreed 
with the proposal in \cite{Beisert:2006ib}
upon zeta function regularization
in all cases tested (all pairs of $r,s$ with $s\leq 10$
to 20 orders). 
In \appref{sec:strong} we will finally show for general $(r,s)$ 
that $c_{r,s}(g)$ at large $g$ yields
the first two orders $c^{(0)}_{r,s}$ and $c^{(1)}_{r,s}$ in \eqref{eq:coeff01}
which are known from perturbative string theory \cite{Arutyunov:2004vx,Hernandez:2006tk}.

\medskip

It is therefore fair to claim that the 
strong-coupling expansion \eqref{eq:cexp.strong}
and the weak-coupling expansion \eqref{eq:cexp.weak}
describe one and the same function $c_{r,s}(g)$ for all $(r,s)$.
However, we should stress that the weak-coupling expansion is more 
versatile than the strong-coupling one:
The functions $h_{r,s}$ have singularities at $g=\pm\frac{i}{4}$,
and thus the coefficients $c_{r,s}(g)$ have singularities
at $g=\pm\frac{i}{4}k$ for all positive integers $k$.%
\footnote{The singularities arise from
the elliptic integrals in \protect\eqref{hKEform}
and therefore appear to be related to a degeneration of the
complex tori defining the kinematical space for the bound states in 
\protect\cite{Dorey:2006dq,Chen:2006gq,Roiban:2006gs,Beisert:2006qh}.}
Consequently, $c_{r,s}(g)$ has a finite radius of convergence
around $g=0$, and the weak-coupling series
\emph{defines} the function
unambiguously. In contradistinction the strong-coupling
expansion is merely asymptotic and does not fully define the function.

\section{Magic Kernels}
\label{sec:kernels}

Now we will return to our study of the integral equation
for the fluctuation density in the presence of a perturbative
dressing phase $\theta \neq 0$. 
As noted in section \ref{sec:dressedES},
there appears to exist a very special choice for the 
constants $\beta_{r,s}^{(\ell)}$ in \eqref{dressing} such
that the odd-zeta contributions to the scaling function
based on a trivial phase $\theta=0$ 
($f\indup{m}(g)$, see \eqref{oldscalingshort}) 
either cancel ($f_{0}(g)$, see \eqref{scalingsym}), or synchronize 
their relative sign w.r.t.~the even-zeta contributions
($f(g)$, see \eqref{scalingshort}). 
The constants of the latter case appeared to
differ from the ones of the former case by a factor of two.
Let us now prove these statements, and derive the 
relevant set of constants $\beta_{r,s}^{(\ell)}$.

An observation of central importance for 
this section is that the 
simplified scaling function $f_{0}(g)$
may be obtained from an effective
kernel much simpler than \eqref{modkernel,dresskernel}.
One just needs to decompose the ``main scattering'' kernel  
\eqref{mainkernel} into two parts
\[
\hat K\indup{m}(t,t')=
\hat K_0(t,t')+
\hat K_1(t,t'),
\]
which are even and odd functions, respectively, under both $t\to -t$, $t'\to -t'$.
Explicitly these functions read
\<\label{symkernel}
\hat K_0(t,  t') \eq
\frac{t\besselJ{1}(t) \besselJ{0}(t')  - t'\besselJ{0}(t) \besselJ{1}(t')}{t^2 -  t'^2}\, ,
\nln
\hat K_1(t, t') \eq
\frac{t' \besselJ{1}(t) \besselJ{0}(t') - t\besselJ{0}(t) \besselJ{1}(t')}{t^2  -  t'^2}\, .
\>
It turns out that the component $\hat K_1$ is responsible
for all odd-zeta contributions to $f\indup{m}$.
If we eliminate it and set $\hat K=\hat K_0$ in \eqref{fredholm}
we obtain just the same functions $\hat\sigma_0$ 
and $f_0(g)$, see \eqref{scalingsym}.
Note that again \eqref{dualpotential} remains true under this 
symmetrization, and \eqref{potential} holds with  
$\hat K\indup{m} \to \hat K_0$. 

\medskip

We can use this observation to obtain an almost closed 
expression for the dressing kernel.
We start from the integral equation with a purely even kernel
\eqref{symkernel}
\[\label{fredholm0}
\hat \sigma_0(t)  =  \frac{t}{e^{t } - 1} 
 \Bigl[  \hat K_0(2\, g \, t,0) -
4\, g^2 \int_0^\infty dt' \; \hat K_0(2\, g \, t, 2\, g \, t') \,
\hat \sigma_0(t') 
\Bigr] ,
\]
which leads to $f_0(g)$.
Replacing $\hat K_0$ by $\hat K\indup{m}-\hat K_1$, 
using $\hat K_1(t,0)=0$, and
substituting $\hat \sigma_0$ once into the ensuing second
convolution only, we derive the relation 
\<
\label{fredholmiter}
\hat \sigma_0(t) \eq  \frac{t}{e^{t } - 1} 
\Bigl[ \hat K\indup{m}(2\, g \, t,0)+ \hat K\indup{c}(2\, g \, t,0)
\nl\qquad\qquad
-4\, g^2 \int_0^\infty dt' \; 
\left( \hat K\indup{m}(2\, g \, t, 2\, g \, t') +
\hat K\indup{c}(2\, g \, t, 2\, g \, t')\right)
\hat \sigma_0(t') 
\Bigr] .
\>
with the function $\hat K\indup{c}$ 
\[\label{crossingkernel}
\hat K\indup{c}(t,t')=4\,g^2\,\int_0^\infty dt''\,
\hat K_1(t,2\,g\,t'')\,\frac{t''}{e^{t''}-1}\,
\hat K_0(2\,g\,t'',t') .
\]

Now \eqref{fredholmiter} is just the original integral
equation \eqref{fredholm}, but with the kernel
$\hat K=\hat K\indup{m}+\hat K\indup{c}$. 
Therefore we can interpret $\hat K\indup{c}$
as a dressing kernel to generate 
the simplified dressing function $f_0(g)$. 
In fact, it agrees \emph{precisely} with the expansion
of the general dressing kernel \eqref{modkernel,dresskernel}
using the first few coefficients found in \eqref{cancelzeta}!
The expression for the dressing kernel $\hat K\indup{c}$ 
allows us to derive a closed expression for
the constants  $\beta_{r,s}^{(\ell)}$
leading to cancellation of
odd-zeta contributions. 
The details of this derivation are presented in \appref{app:magicexpand},
the final result is 
\[
\label{naivecoeffs}
\beta_{r,r+1+2\nu}^{(r+\nu+\mu)} \to
(-1)^{r+\mu+1}\,\frac{(r-1)(r+2\nu)}{2\mu+1}\,
\left(\atopfrac{2\mu+1}{\mu-r-\nu+1}\right)
\left(\atopfrac{2\mu+1}{\mu-\nu}\right)
\zeta(2\mu+1)\, .
\]

Surprisingly, these coefficients agree, up to a global factor of $\half$, 
with the ``analytic continuation'' of the conjectured 
strong-coupling dressing phase constants \eqref{eq:WL,result}!
The results of the previous section then suggest that the correct
crossing-invariant choice of constants is just twice \eqref{naivecoeffs}.
Of course, these constants agree with those in \eqref{flipsigns}.
The corresponding kernel for the integral equation \eqref{fredholm} of
the universal scaling function $f(g)$ is,
cf.~\eqref{mainkernel,crossingkernel}
\[
\hat K(t,t')=\hat K\indup{m}(t,t')+2\,\hat K\indup{c}(t,t')\, .
\]

\medskip

What does this imply for the universal scaling function?
We ``experimentally'' observed in section \ref{sec:dressedES}, to 
rather high order, that the doubled constants \eqref{truecoeffs}
lead to sign-reversals in those odd-zeta terms which do not
have the same relative sign as the even-zeta terms in the
original proposal \eqref{oldscalingshort}. 
Up to this minor modification,
the associated scaling function agrees with the one based
on a trivial phase. 
We will present a proof of this statement in \appref{app:flipzeta}.

It is interesting to investigate the analytic structure of the 
three candidate scaling functions. 
In each case one would like to find the locations 
of all singularities in the complex $g$-plane,
and determine wether they correspond to poles,
branch points, or are of essential type. 
The convergence properties of the weak coupling expansion 
series  are, in particular, clearly related to the singularity with 
smallest $|g|$.  

\begin{figure}\centering
\includegraphics[width=7cm]{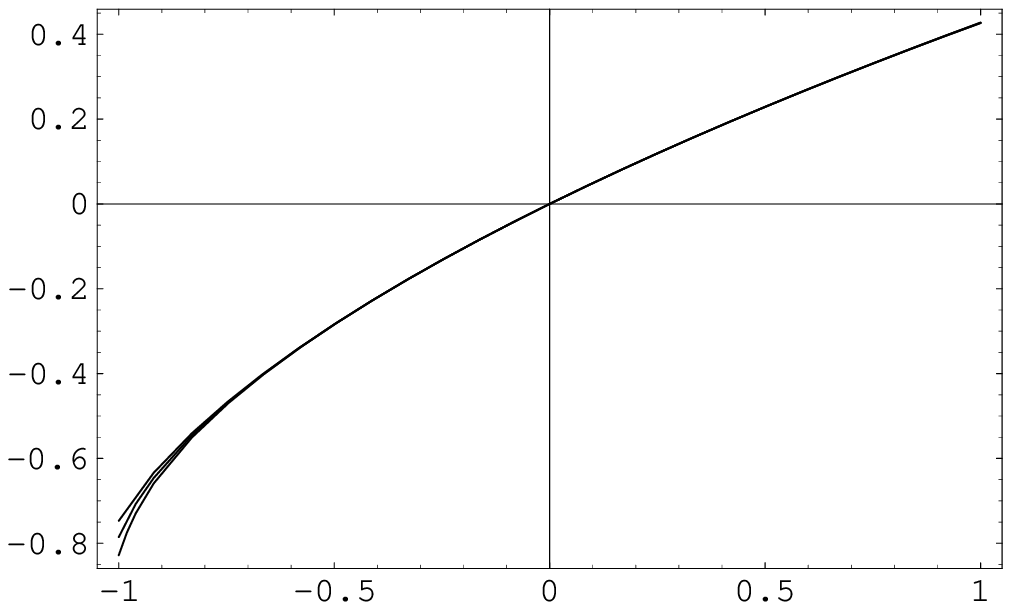}
\caption{
Top to bottom curve: 
Plot of the scaling functions $f\indup{m}(g)$, $f_0(g)$ and $f(g)$ with $g=\quarter\sqrt{x}$.}
\label{fig:cuspdim}
\end{figure}

The integral form of the dressing kernel $K\indup{c}$ allows
to derive conveniently the first few orders 
in the expansion of the scaling function.
The analytic expression is given \eqref{scalingshort}.
Beyond $20$ orders, however, the analytic calculation
slowed down substantially due to an exploding number
of partitions for the odd-zeta contributions. 
This complexity can be reduced by reverting 
to a numerical computation. We were able to produce
all three scaling functions $f\indup{m}(g)$, $f_0(g)$ and $f(g)$
up to $50$ loops, $\order{g^{100}}$, at $1000$-digit accuracy 
(needed for intermediate steps).
A plot of these three functions is shown in \figref{fig:cuspdim}.

We have then used the quotient criterion 
to investigate the convergence properties of the series.
Our results indicate that the radius of convergence in $g$
is $\quarter$ in all three cases. 
The related singularity is situated at $g=\pm \frac{i}{4}$
or, put differently, at $\lambda=-\pi^2$.
The accuracy of this result was very high
and the position matches with the singularity of
the function in \eqref{eq:ch23}.
The exponents of the singularity
$f\simeq (\lambda+\pi^2)^\alpha$ appear
to be $\alpha=+1,+\frac{2}{3},+\frac{1}{2}$ for the
three functions $f\indup{m}(g)$, $f_0(g)$, $f(g)$, respectively.
We were not able to produce a very accurate result in 
this case, we expect it to be accurate within a few percent. 

\medskip

In conclusion, the final proposal for the, hopefully correct, 
expansion coefficients of the dressing phase 
\eqref{dressing,betaexpand} reads
\[
\label{truecoeffs}
\beta_{r,r+1+2\nu}^{(r+\nu+\mu)} =
2 (-1)^{r+\mu+1}\,\frac{(r-1)(r+2\nu)}{2\mu+1}\,
\left(\atopfrac{2\mu+1}{\mu-r-\nu+1}\right)
\left(\atopfrac{2\mu+1}{\mu-\nu}\right)
\zeta(2\mu+1)\, .
\]
The resulting universal scaling function $f(g)$ 
in a weak-coupling expansion was presented in \eqref{scalingshort}
to the first few orders.
Its convergence properties seem to indicate a singularity
at $\lambda=-\pi^2$ with exponent $+\frac{1}{2}$.

\medskip

The strong-coupling expansion of 
the scaling function $f(g)$ remains to be evaluated
within the Bethe ansatz framework. 
A comparison to available data \cite{Gubser:2002tv,Frolov:2002av}
from string theory would constitute an important test 
of our proposed dressing phase. 
The leading order at strong coupling 
based on the AFS phase \cite{Arutyunov:2004vx}
(which is consistent with our proposal)
indeed agrees \cite{Frolov:2006aa} with an explicit 
string theory calculation \cite{Gubser:2002tv}.
Conversely, in the case of the un-dressed function $f\indup{m}(g)$, 
the integral equation is troublesome and leads
to a fluctuating solution \cite{Lipatov:2006aa}. Hopefully
the situation is improved by the correct dressing phase
in order to find agreement with string theory.

\section{Conclusions and Outlook}
\label{sec:concl}

In this paper we have derived an expression 
\eqref{dressing,betaexpand,truecoeffs}
for the complete weak-coupling expansion of the 
dressing phase $\theta$ in 
the Bethe ansatz \cite{Beisert:2005fw} 
of planar $\superN=4$ gauge theory. 
It is based on a crossing-symmetric
proposal in Sec.~5.2 of \cite{Beisert:2006ib} for the 
strong-coupling expansion of the dressing phase
in conjunction with some curious structures of cancellations 
and replacements when a dressing phase modifies
the scaling function $f\indup{m}(g)$ based
on a trivial phase, as originally derived in \cite{Eden:2006rx}.
Whether or not our proposal describes the AdS/CFT system
accurately, we have shown that there exists a 
natural expression for the dressing phase which 
appears to interpolate between 
currently known spectral data at 
weak and strong coupling.
We are hopeful, but have not shown, that any alteration
of our proposal is likely to violate the transcendentality principle
in gauge theory and/or crossing symmetry in string theory as well as
leads to potential contradictions with the available data. 
In that sense our proposal lends support to
an exact interpretation of the AdS/CFT correspondence
\cite{Maldacena:1998re,Gubser:1998bc,Witten:1998qj}, at least in the case
of $\superN=4$ gauge theory in the 't~Hooft limit
and free IIB superstrings on $AdS_5\times S^5$.

The dressing phase $\theta$ is given as a multiple expansion in
excitation charges and the coupling constant. The radius of
convergence w.r.t.~the 't Hooft coupling $\lambda$ 
around $\lambda=0$ is $|\lambda|<\pi^2$.
The singularities are situated at $\lambda=-n^2\pi^2$ 
for all non-zero integers $n$. 
Consequently, perturbative gauge theory appears to unambiguously define 
the phase and its analytic continuation. 
Note however that we have not succeeded in summing over the modes $r,s$ of
$\beta_{r,s}(g)$ or even study the analytic structure of the phase 
$\theta$ in general.
Conversely, the strong-coupling
expansion of the phase can merely be asymptotic 
because $\lambda=\infty$ is an accumulation point of singularities.
The asymptotic string expansion proposed in Sec.~5.2 of \cite{Beisert:2006ib}
seems to be in perfect agreement with our proposal, 
but it is not sufficient to \emph{uniquely} define a complex function.

\medskip

Two methods of testing our coefficients come to mind: 
The first is a direct computation of the 
scaling function $f(g)$ at four loops.
Until recently, it was known up to three loops
\cite{Anastasiou:2003kj,Bern:2005iz},
see also \cite{Kotikov:2004er},
but with methods based on the unitarity of scattering amplitudes
it is within computational reach
\cite{Bern:2006aa,Kosower:2006aa,Bern:2006ew}.
Our prediction $\beta^{(3)}_{2,3}=4\beta=4\zeta(3)$ leads
to a four-loop contribution to the universal scaling function
\<
f(g)\eq\ldots
- 16\,\left( \frac{73}{630}\,\pi^6+4\,\zeta(3)^2 
\right)\, g^8 \,+\ldots
\nln
\earel{\approx}
\ldots-
3.01502 \times 10^{-6} \lambda^4+\ldots\,.
\>
The final outcome of the four-loop calculation
by Bern, Czakon, Dixon, Kosower and Smirnov 
\cite{Bern:2006ew},
which was performed in parallel and completely 
independently from our analysis, is
\<
f(g)\eq\ldots
- 64\times\left( 29.335\pm0.052 \right)\, g^8 \,+\ldots
\nln
\eq
\ldots-
\bigbrk{3.0192\pm 0.0054}
\times 10^{-6} \lambda^4+\ldots\,.
\>
The perfect agreement between the two numbers 
provides further confidence that our proposed 
dressing phase correctly describes the asymptotic spectrum
of anomalous dimensions in
planar $\superN=4$ gauge theory. 
At the least it ascertains that the dressing phase is non-trivial
starting from $\order{g^6}$, and that
the constant for the first correction in \cite{Eden:2006rx}
takes the value $\beta=\quarter\beta^{(3)}_{2,3}=\zeta(3)$.
Note, however, that we have to assume the exact universality of the
scaling function, i.e.~that it is exactly the same for twist-two
(as assumed in \cite{Bern:2006ew}) and for higher-twist
(as required for the asymptotic Bethe ansatz).

A computation of the function $f(g)$ to even higher loop orders 
would provide more stringent tests of our proposal. 
Given the remarkable success of unitarity-based methods
in \cite{Anastasiou:2003kj,Bern:2005iz,Bern:2006ew}
we do not dare calling this impracticable.
However, some yet-to-be-discovered iterative structure is likely to exist 
if our comparably simple integral equation holds true.
With many perturbative orders available
an extrapolation to strong coupling may be attempted
and compared to string theory results \cite{Gubser:2002tv,Frolov:2002av}
via the AdS/CFT correspondence.
Such an analysis was indeed performed in \cite{Bern:2006ew}
using the KLV \cite{Kotikov:2003fb} and Pad\'e approximations.
It is based on their independent guess 
of the function $f(g)$ in \eqref{scalingshort}
and how to obtain it from the earlier proposal \eqref{oldscalingshort}
by synchronizing all signs.
Strangely, their approximations, 
based on using the first seven or eight terms for example, 
appear to indicate that \eqref{scalingshort} is about $5\%$ away 
from the expected strong-coupling value of \cite{Gubser:2002tv}.
Non-coincidence of the values would be rather
puzzling with regard to the discussion at
the end of \secref{sec:kernels} and \secref{sec:analytic}.
Would it imply that $f(g)$ at strong coupling
does not describe the string states in \cite{Gubser:2002tv,Frolov:2002av}?
Are there some non-perturbative corrections?
To that end, it would be worth investigating whether an adapted scheme 
which takes into account all singularities at $\lambda=-n^2\pi^2$ 
and their accumulation at $\lambda=\infty$ can produce 
an extrapolation of $f(g)$ closer to 
the expected string theory values \cite{Gubser:2002tv,Frolov:2002av}.

\medskip

Another way of testing the proposal involves 
local operators of small classical dimension.
A gauge Bethe ansatz modified 
by a dressing phase changes the anomalous 
dimensions $\Delta$ of all operators in all sectors. 
E.g.~in the $\alg{su}(2)$ sector we would find for the length $L=5$ 
operator $\Tr X^2\,Z^3 + \ldots$ (this case is actually equivalent
to the $\alg{sl}(2)$ twist=length three operator $\Tr D^2\,Z^3 + \ldots$)
to four loops
\[
\Delta=5+8\,g^2-24\,g^4+136\,g^6-16\,\bigbrk{\sfrac{115}{2}+2\,\beta_{2,3}^{(3)}}\,g^8\, + \ldots \, .
\]
For the specific case of constants \eqref{truecoeffs} we then have
\[
\label{L5prediction}
\Delta=5+8\,g^2-24\,g^4+136\,g^6
-16\,\bigbrk{\sfrac{115}{2}+8\,\zeta(3)}\,g^8\, + \ldots \, .
\]
Intriguingly, this predicts the appearance of a non-rational term in the
four-loop anomalous dimension of a finite length  operator! 
Such contributions are impossible if the dressing phase is trivial, i.e.~if $\theta=0$. 
They are certainly not excluded, and a priori even likely, from the point
of view of perturbative field theory.
This would then also rule out the all-loop ``BDS'' ansatz originally 
proposed in \cite{Beisert:2004hm}, and as a consequence also its
finite-length description through the Lieb-Wu equations
of the Hubbard model, cf.~(68) in \cite{Rej:2005qt}.
It also eliminates the Hubbard Hamiltonian as an exact
candidate for the $\alg{su}(2)$ dilatation operator \cite{Rej:2005qt}.
However, a gauge theory dressing phase \eqref{dressing} whose constants preserve
transcendentality for twist operators, such as in the specific suggestion \eqref{truecoeffs},
would have the following intriguing property affecting operators of finite length and spin:
Up to (at least) wrapping order, the BDS ansatz would give the correct ``rational part''
of all anomalous dimension matrices, 
and the Hubbard Hamiltonian would emulate the ``rational part''
of the $\alg{su}(2)$ dilatation operator to (at least) 
wrapping order.

An interesting question concerns the computation of
anomalous dimensions beyond wrapping order. 
The simplest case is the four-loop anomalous dimension
of the Konishi field, i.e. the $\alg{su}(2)$  length $L=4$ 
operator $\Tr X^2\,Z^2 + \ldots$, or, equivalently, 
in $\alg{sl}(2)$ the lowest spin $S=2$, length $L=2$
operator $\Tr D^2\,Z^2 + \ldots$.
If we, a priori ``illegally'', apply the asymptotic Bethe ansatz
\eqref{allloopbethemod,dressing} we find the four-loop 
result
\[
\Delta=4+12\,g^2-48\,g^4+336\,g^6
-16\,\bigbrk{\sfrac{705}{4}+\sfrac{9}{2}\,\beta_{2,3}^{(3)}}\,g^8\, + \ldots \, .
\]
Recall that $\beta_{2,3}^{(3)}=0$ is the ``BDS'' case with
dressing phase $\theta=0$. With our conjecture
\eqref{truecoeffs} we then have instead
\[
\Delta=4+12\,g^2-48\,g^4+336\,g^6
-16\,\bigbrk{\sfrac{705}{4}+18\,\zeta(3)}\,g^8\, + \ldots \, .
\]
However, as opposed to \eqref{L5prediction}, this result
is much less certain. It would only be true if our
asymptotic Bethe equations turn out to be exact.
Conversely, recall that the Hubbard model 
\cite{Rej:2005qt} yields for this state
\[
\Delta=4+12\,g^2-48\,g^4+336\,g^6-16\times318\,g^8 + \ldots \, .
\]
Could it be that the Hubbard model still properly
yields the ``rational part'' of fields even beyond
wrapping order? This might then result in 
\[
\Delta=4+12\,g^2-48\,g^4+336\,g^6
-16\,\bigbrk{318+18\,\zeta(3)}\,g^8\, + \ldots \, ,
\]
but, in the absence of a yet-to-be constructed rigorous Bethe 
ansatz for the finite system, this is of course a quite unfounded 
speculation.

\medskip

Of course, there are more general questions:
It would be very valuable to find a closed form 
for the proposed dressing phase $\theta$
and to prove its correctness. Likewise, the exact integrability
of the planar AdS/CFT model still lacks a rigorous proof. 
Are the Bethe equations at finite coupling \emph{and} finite length 
exact or is there a more fundamental description of the planar spectrum 
of which the Bethe equations are merely some limits?
For instance, the exactness may be flawed by certain terms 
which are exponentially suppressed 
at long states \cite{Schafer-Nameki:2005tn,Schafer-Nameki:2006ey}.
The ``covariant'' approaches started in 
\cite{Mann:2005ab,Rej:2005qt,Gromov:2006dh,Gromov:2006cq}
may provide an answer to this question as well as 
recover the dressing phase
from a more fundamental (and simpler) description.
Other promising approaches
are the thermodynamical Bethe ansatz
(see \cite{Ambjorn:2005wa})
and Baxter equations (see \cite{Teschner:2006aa,Bytsko:2006ut}).

\medskip

It is curious to see that in our proposal the deviations from 
a trivial phase start to contribute to anomalous dimensions 
at four loops, which is a rather high perturbative order.
If true, this might indicate that intuition gained from
some perturbative computations,
even if they extend to two or three loops,
may be deceiving. 
For instance, a non-zero value for $\beta^{(3)}_{2,3}=4\beta$ 
leads to a four-loop breakdown 
of weak-coupling BMN-scaling \cite{Berenstein:2002jq} in
$\superN=4$ gauge theory.
This possibility was discussed by several authors in
the past in order to reconcile discrepancies between 
gauge and string theory structures at, respectively weak and 
strong coupling. See in particular
\cite{Klebanov:2002mp,Serban:2004jf,Beisert:2004hm,Beisert:2004yq,Beisert:2005cw,Klebanov:2006aa,Staudacher:2006aa}.
Our proposal suggests that this is indeed the case.
The assumption of \cite{Berenstein:2002jq}
that the BMN limit leads to a dilute gas with
weakly interacting particles
is not compatible with our phase.
It should be noted that the perturbative 
dispersion relation obtained in \cite{Gross:2002su,Santambrogio:2002sb}
is indeed compatible with our proposal. 
However, it cannot be translated directly into 
anomalous dimensions of local operators
as suggested in \cite{Berenstein:2002jq,Gross:2002su,Santambrogio:2002sb}.

The same applies for the spinning strings proposal
\cite{Frolov:2003qc}:
The weak-coupling expansion in the effective coupling
constant $\lambda'\sim\lambda/J^2$ 
leads to terms of the type $(\lambda')^4 J^2$ which 
diverge in the limit of large spin $J$.
The investigation of spinning string solutions
from the gauge theory side started in 
\cite{Beisert:2003xu} will therefore 
fatally break down at the four-loop level.
Nevertheless, the proposals 
\cite{Berenstein:2002jq,Frolov:2003qc}
have been of tremendous importance 
to initiate the study of stringy
and integrable structures in $\superN=4$ gauge theory.

Finally, it is worth pointing out
that $\beta_{2,3}(g)=\order{g^6}$
is reminiscent of the findings 
\cite{Fischbacher:2004iu}
within the plane wave (BMN) matrix model \cite{Berenstein:2002jq}.
Also there, a non-trivial dressing phase was observed starting
at the same order. The main difference is that the coefficient
$\beta_{2,3}^{(3)}$ for the BMN matrix model is rational while
ours is a rational multiple of $\zeta(3)$. This may potentially
be related to the fact that the matrix model has only finitely
many degrees of freedom while $\superN=4$ SYM is non-compact.

\subsection*{Acknowledgements}

We are grateful to
Z.~Bern,
M.~Czakon,
L.~Dixon,
D.~Kosower and
V.~Smirnov
for supplying us with a draft of their manuscript
\cite{Bern:2006ew}, and for discussions as well as comments
on our draft. 
We would also like to sincerely thank them for undertaking such a 
tremendously important and impressive effort.
We furthermore thank S.~Frolov, E.~L\'opez, J.~Plefka and D.~Serban 
for comments on the manuscript.

\appendix

\section{Expansion of the Magic Kernel}
\label{app:magicexpand}

In this appendix we present a weak-coupling expansion 
of the magic kernel $\hat K\indup{c}$ 
in \eqref{crossingkernel}
leading to 
an expression for the coefficients $\beta^{(\ell)}_{r,s}$
of the dressing phase.

A useful integral representation for the even and odd parts was found in 
\cite{Eden:2006rx}:
 \<\label{k01int}
\hat K_0(t,t') \eq \int_0^1 d\lambda \, \lambda\,\besselJ{0}(\lambda\, t) \besselJ{0}(\lambda\, t') \,, 
\nln
\hat K_1(t,t') \eq \int_0^1 d\lambda \, \lambda\,\besselJ{1}(\lambda\, t) \besselJ{1}(\lambda\, t') \, .
\>
With their help, we may calculate $\hat K\indup{c}$ to be
\< \nonumber
\hat K\indup{c}(t,t') \eq
4 g^2 \int_0^\infty dt'' \; \int_0^1 d\lambda_1 \,
\lambda_1 \besselJ{1}(\lambda_1\, t) \besselJ{1}(\lambda_1\, 2\, g \, t'') \;
\frac{t''}{e^{t''} \, - \, 1} \, \times 
\nl 
\phantom{2 g^2\int_0^\infty dt'' \;} \int_0^1 d\lambda_2 \, \lambda_2 \,
\besselJ{0}(\lambda_2\, 2\, g \, t'')  \besselJ{0}(\lambda_2\, t')
\nln
\eq - 4 \sum_{\mu=1}^\infty (-1)^\mu
g^{2 \mu+1} \zeta(2 \mu+1) \; (2
\mu)! \, \times 
\nl\qquad
 \sum_{n=0}^{\mu-1} \frac{1}{n! \,
(n+1)! \, (\mu-1-n)! \, (\mu-1-n)!} \, \times 
\nl\qquad
\int_0^1
d\lambda_1 \, \lambda_1^{2(n+1)}  \besselJ{1}(\lambda_1\, t) \, \int_0^1
d\lambda_2 \, \lambda_2^{2(\mu-n)-1}  \besselJ{0}(\lambda_2\, t')
\label{lastI}
\>
where we have first expanded the Bessel functions with argument
$\lambda_{1,2} \,2\, g \, t''$ in $g$ and then executed the $t''$
integration. The two remaining parameter integrals in the last
line may be found from a standard formula. In the first integral
the power of $\lambda_1$ in front of $\besselJ{1}(\lambda_1 t)$ is always
even, and in the second one the power of $\lambda_2$ in front of
$\besselJ{0}(\lambda_2 t')$ is always odd. Both integrals therefore yield
finite sums of Bessel functions:
\<
\earel{} \int_0^1 d\lambda_1 \, \lambda_1^{2 m}  \besselJ{1}(\lambda_1\, t) 
\nln\eq - \frac{(m-1)! \, m!}{(2 m)!} \sum_{k=1}^m (-1)^k
\begin{pmatrix} 2 m \\ m-k \end{pmatrix} \frac{2 k}{t}  \besselJ{2 k}(t)
\nln\earel{}
\int_0^1 d\lambda_2 \, \lambda_2^{2 m-1} \besselJ{0}(\lambda_2\, t')
\nln\eq
 - \frac{(m-1)! \, (m-1)!}{(2 m -1)!}
\sum_{l=1}^m (-1)^l\begin{pmatrix} 2 m -1 \\ m-l
\end{pmatrix} \frac{2 l-1}{t'} \, \besselJ{2 l-1}(t')\, . 
\>
On substituting these into \eqref{lastI} we can cancel the factors
which depend on $n$ but not on $k,l$. We re-order the $n,k,l$ sums
to find:
\<
\hat K\indup{c}(t,t') \eq - \frac{4}{t \, t'} \sum_{\mu=1}^\infty (-1)^\mu
g^{2 \mu+1} \zeta(2 \mu+1) \; (2
\mu)! \, \times 
\nl\qquad
 \sum_{k,l \geq 1}^{k+l \leq \mu + 1} (-1)^{k+l}(2 k) \besselJ{2 k}(t) \; (2 l - 1) \besselJ{2 l - 1}(t') \, \times 
\nl\qquad
\sum_{n = k-1}^{\mu-l} \frac{1}{(n+1-k)! \, (n+1+k)! \, (\mu-l-n)!
\, (\mu+l-1-n)!} \, .
\>
The sum over $n$ can be done in closed form:
\<
\earel{} \sum_{n = k-1}^{\mu-l} \frac{1}{(n+1-k)! \, (n+1+k)! \,
(\mu-l-n)! \, (\mu+l-1-n)!} 
\nln\eq
\frac{1}{(2 \mu
+1)!} \begin{pmatrix} 2 \mu + 1 \\ \mu + 1 - k - l \end{pmatrix}
\begin{pmatrix} 2 \mu + 1 \\ \mu + 1 + k - l \end{pmatrix}\, .
\>
Collecting results, we get
\<
\hat K\indup{c}(t,t')\eq - \frac{4}{t \, t'} \sum_{\mu=1}^\infty
g^{2 \mu+1} \, \sum_{k,l \geq 1}^{k+l \leq \mu + 1}  \besselJ{2 k}(t)  \besselJ{2 l - 1}(t') \,
(-1)^{\mu+k+l} \, \times 
\nl\qquad
\frac{(2 k) \, (2 l - 1)}{2
\mu + 1} \begin{pmatrix} 2 \mu + 1 \\ \mu + 1 - k - l
\end{pmatrix} \begin{pmatrix} 2 \mu + 1 \\ \mu + 1 + k - l \end{pmatrix}
\zeta(2 \mu+1)\, . \label{solvedIt}
\>
This is \emph{precisely} of the form of a kernel stemming from a non-trivial
dressing phase $\theta$, as we can see by comparing to
\eqref{dresskernel}! 
We can even read off the constants $\beta_{r,s}^{(\ell)}$:
\[
\beta_{r,r+1+2\nu}^{(r+\nu+\mu)} \to
(-1)^{r+\mu+1}\,\frac{(r-1)(r+2\nu)}{2\mu+1}\,
\left(\atopfrac{2\mu+1}{\mu-r-\nu+1}\right)
\left(\atopfrac{2\mu+1}{\mu-\nu}\right)
\zeta(2\mu+1)\, .
\]

\medskip

Finally, instead of using the integral representation \eqref{k01int}, we may
also cast the two kernels $\hat K_0$, $\hat K_1$ in the form of infinite sums:%
\footnote{This result was first obtained by D.~Serban \protect\cite{Serban:2006aa}.}
\<
\hat K_0(t,t') \eq \frac{2}{t \, t'} \sum_{n=0}^\infty (2 n+1) \besselJ{2 n+1}(t)  \besselJ{2 n+1}(t') \, , 
\nln
\hat K_1(t,t') \eq \frac{2}{t \, t'} \sum_{n=1}^\infty (2 n) \besselJ{2 n}(t)  \besselJ{2 n}(t') \, . \nonumber
\>
It follows that
\[
\hat K\indup{c}(t,t') = 
\frac{2}{t \, t'} 
\sum_{k=1}^\infty 
\sum_{l=0}^\infty
(-1)^{k+l} c_{2k+1,2l+2}(g)  \besselJ{2 k}(t)  \besselJ{2 l + 1}(t') 
\]
with the coefficient functions
\[
\label{cintrep}
c_{r,s}(g)=2\cos[\half\pi(s-r-1)]\,(r-1)(s-1)
\int_0^\infty dt\, \frac{\besselJ{r-1}(2\,g\,t)\besselJ{s-1}(2\,g\,t)}{t(e^t-1)}\,.
\]
We can use this form to present an alternative proof.
We firstly expand $1/(e^t-1)$ into a geometric series
and secondly rescale the integration variable $t$ 
by $g$ such that the integral will manifestly depend on 
the ratio $g/n$ only
\<
\label{cintrepsum}
c_{r,s}(g)\eq 
\sum_{n=1}^\infty h_{r,s}(g/n),
\\\nn
h_{r,s}(g/n)\eq 2\cos[\half\pi(s-r-1)]\,(r-1)(s-1)
\int_0^\infty \frac{dt}{t}\,\exp(-tn/g) 
\besselJ{r-1}(2\, t)\besselJ{s-1}(2\, t)\, .
\>
The integral can now be performed and yields precisely 
the hypergeometric function derived earlier in \eqref{hhyper}.

\medskip

The discussion of the strong-coupling behavior of the dressing phase is
tantamount to an analysis of the strong-coupling limit of the integrals $c_{r,s}(g)$. 
In \appref{sec:strong} we present, by way of example,
the full expansion in the case $(r,s)=(2,3)$. We also show
that \eqref{cintrepsum} reproduces the AFS phase \cite{Arutyunov:2004vx}
as well as the Hern\'andez-L\'opez correction \cite{Hernandez:2006tk}.

\section{Flipping Odd-Zetas Contributions}
\label{app:flipzeta}

Here we show that the replacement of the
kernel 
$\hat K\indup{m}\to\hat K\indup{m}+2\hat K\indup{c}$
leads to 
the simple replacement $\zeta(2n+1)\to i\zeta(2n+1)$ in 
the scaling function $f\indup{m}(g)\to f(g)$.

Let
\[
z(t)=\frac{t}{e^t-1}\,.
\]
For any kernel $\hat K(2 g t,  2 g t')$ and any function $f(t)$ we define the
product
\[
(\hat K \mathop{\ast} f)(t) = \int_0^\infty dt' \, \hat K(2\, g\, t,  2\, g\, t') \,
z(t') \, f(t') \, . \label{defstar}
\]
The integral equation may be written as
\[
\hat\sigma(t)  = z(t) \left( (\hat K_0 + \hat K\indup{c})(2\, g\, t,0)  -  4 g^2 \,
\left( (\hat K_0 + \hat K_1 + \hat K\indup{c}) \mathop{\ast} \frac{\hat\sigma}{z} \right)(t) \right)
\, . \label{symbolicFred}
\]
In the following we need not consider the internal $g$-dependence of the
kernels. Within the radius of convergence of the iteration the
solution of any Fredholm equation of the second type is the resolvent
\[
\hat\sigma(t) = z(t) \sum_{n=0}^\infty \, (-4 g^2)^n \, 
\biggl( [ (\hat K_0 + \hat K_1 + \hat K\indup{c}) \mathop{\ast} ]^n \, P \biggr)(t) \, .
\]
Here $P(t) = (\hat K_0 + \hat K\indup{c})(2 g t, 0)$ and $\hat K\indup{c}(2 g t,  2 g t') = 8 g^2
(\hat K_1 \mathop{\ast} \hat K_0)(2 g t, \, 2 g t')$.
The Fredholm resolvent must have the form
\[
\hat\sigma(t)  =  z(t) \sum_{\underline{n}} \, (-4g^2)^n \,
\eta_{\underline{n}} \, (W_{\underline{n}} \, \hat K_0)(t) 
\label{fredResolve}
\]
because the dressing kernel $\hat K\indup{c}$ consists itself 
of $\hat K_0,\hat K_1$ which are convoluted by the star operation 
(and it also carries the appropriate extra power of the coupling constant).
By $\underline{n}$ we denote a string built of $0,1$ with total length
$n$, and $W_{\underline{n}}$ is the corresponding string of
$\hat K_0,\hat K_1$ acting on the potential $\hat K_0$ by the $\ast$ product defined above in
\eqref{defstar}. The $\eta_{\underline{n}}$ are numerical coefficients.
The integral equation \eqref{symbolicFred} determines these numbers
uniquely: We substitute \eqref{fredResolve} into the equation and pick the
$\order{g^{2(n+2)}}$ terms. All words are assumed independent; in
particular the four double iterations of an arbitrary word
$W_{\underline{n}}$:
\[
W_{0,0,\underline{n}} \, , \qquad W_{0,1,\underline{n}} \, ,
\qquad W_{1,0,\underline{n}} \, , \qquad W_{1,1,\underline{n}}\,.
\]
This yields the four equations
\begin{eqnarray}
\eta_{0,0,\underline{n}} \eq - \, \eta_{0,\underline{n}} \, ,
\label{allfour}
\nln
 \eta_{0,1,\underline{n}}
\eq 
- \eta_{1,\underline{n}} \, ,
\nln
 \eta_{1,0,\underline{n}}
\eq 
- \eta_{0,\underline{n}} - 2 \eta_{\underline{n}} \, , 
\nln
 \eta_{1,1,\underline{n}} \eq - \eta_{1,\underline{n}} \, . \nonumber
\end{eqnarray}
The first two equations may be summarized as
\[
\eta_{0,\underline{m}}  =  -  \eta_{\underline{m}}
\]
from which it follows that the third equation is:
\[
\eta_{1,0,\underline{n}}  =  +  \eta_{0,\underline{n}}.
\label{dresschange}
\]
Recursively, these equations imply that all the $\eta$
coefficients are merely signs. Second, \eqref{dresschange} implies
that the operation of appending $\hat K_1$ after $\hat K_0$ is the only one
that induces a sign flip as opposed to the iteration of the main scattering
kernel $\hat K\indup{m} = \hat K_0 + \hat K_1$ alone 
(in that case the second term in the r.h.s.
of \eqref{allfour} would be absent so that 
$\eta_{1,\underline{m}}  = -\eta_{\underline{m}}$, too).

Next, let
\[
\mathcal{E}  =  \{ t^{2 l} \, : \; l \in \mathbb{N}_0 \}, \qquad
\mathcal{O}  =  \{ t^{2 l+1} \, : \; l \in \mathbb{N}_0 \},
\]
denote the even and odd powers of $t$, respectively. In order to obtain
the scaling functions at weak coupling we will eventually expand the kernels
$\hat K_{0,1}(2 g t,2 g t')$ in the coupling constant; $\hat K_0$ ($\hat K_1$) is even
(odd) in both arguments, respectively. Let us consider any one term in the
expansion of a kernel convoluted on a power of $t$ by the star product. We
observe the graded structure
\[
\begin{array}{lllllll}
\hat K_0 \, : &\mathcal{E} \earel{\rightarrow} \mathcal{E} & : \qquad t^{2 l}    \earel{\mapsto}  \zeta(2 m)     \, t^{2 l'}, \\ 
              &\mathcal{O} \earel{\rightarrow} \mathcal{E} & : \qquad t^{2 l+1}  \earel{\mapsto}  \zeta(2 m + 1) \, t^{2 l'}, \\ 
\hat K_1 \, : &\mathcal{E} \earel{\rightarrow} \mathcal{O} & : \qquad t^{2 l}    \earel{\mapsto}  \zeta(2 m + 1) \, t^{2 l'+1}, \\ 
              &\mathcal{O} \earel{\rightarrow} \mathcal{O} & : \qquad t^{2 l+1}  \earel{\mapsto}  \zeta(2 m)     \, t^{2 l'+1}.
\end{array}
\]
The first integral in a chain of convolutions always acts on an element of
${\cal E}$ because the potential is even. Likewise, the energy integral
adds a final convolution on the even kernel $\hat K_0(0,2 g t)$. The contribution
of a word $W_{\underline{n}}$ to the scaling function thus contains one odd
zeta function for each beginning and each end of a $\hat K_1$ string
in $\hat K_0 \mathop{\ast} W_{\underline{n}} \mathop{\ast} \hat K_0$. In particular, if there are
$m$ such strings, the word gives a contribution with $2 m$ odd
zeta functions.

Finally, the equations \eqref{allfour,dresschange} say
that each creation of a $\hat K_1$ string induces a sign change as opposed to the
iteration of the main scattering kernel on its own.
Our result is: The scaling functions for both cases contain
terms with $2 m$ odd zeta functions. The relative sign of such
terms is $(-1)^m$.

Note that the absolute sign of the terms with odd zeta functions cannot
easily be fixed without explicit computation. Words with $n$ insertions of
$\hat K_1$ pick up a factor $(-1)^n$ from the $g^2$ expansion relative to those with
$\hat K_0$ alone. Presumably, words with the minimal number of $\hat K_1$ kernels give
the dominating contribution: $\hat K_1$ starts at $\order{g^2}$ thereby
restricting the number of ways one may choose the powers in the various
kernels in a chain, that is reducing the number of similar terms. If this is
so, the iteration of the main scattering kernel should indeed produce
a factor of $(-1)^m$ for a term with $2 m$ odd zeta functions relative to
the leading purely even part. Consequently, the dressing kernel ought to align
all signs.

\section{Strong-Coupling Expansion}
\label{sec:strong}

In this appendix we give an exact derivation of the strong coupling
expansion of $c_{2,3}(g)$. We modify a technique developed in
\cite{Economou:1979aa} for the discussion of the ground state energy of the
half-filled Hubbard model.

\begin{figure}\centering
\includegraphics[scale=0.8]{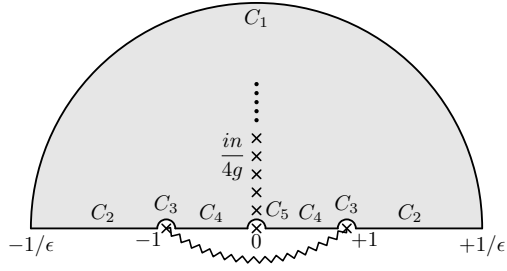}
\caption{Contour $C$ and its decomposition 
\protect\eqref{eq:contour}.}
\label{fig:contour}
\end{figure}

According to equation \eqref{cintrepsum},
\<\label{expandE}
c_{2,3}(g) \eq 
4\sum_{n=1}^\infty \int _0^\infty dy\, 
\frac{\besselJ{1}(2\, g\, y) \besselJ{2}(2\, g\, y)}{y} \, e^{- n y} 
=
\sum_{n=1}^\infty if'\lrbrk{\frac{in}{4g}}
\>
with
\<
f'(z)\eq 4z -4z \, \hypergeofn{2}{1}(-\half,\half;2;z^{-2})
-\frac{1}{2z} \, \hypergeofn{2}{1}(\half,\sfrac{3}{2};3;z^{-2} ),
\nln
f(z)\eq
2 z^2 
\lrbrk{1-\hypergeofn{2}{1}(-\half,\half;2;z_n^{-2} ) }.
\>
We want to use the residue theorem to express the sum as a contour integral.
For any holomorphic function $f(z)$
\[
\text{res}_{in/4 g} \, \frac{f(z)}{\sinh^2(4 \pi g z)}  
= \frac{1}{(4 \pi g)^2} \, f'\left(\frac{in}{4 g}\right)  
\]
because the terms in which the derivative does not fall on $f(z)$
cancel out.%
\footnote{The original article \cite{Economou:1979aa} uses a
simple $\sinh$ in the denominator. This would lead to an alternating
sum of residues, though.} 
The integral is thus written as
\[
c_{2,3}(g)
=\frac{(4 \pi g)^2}{2 \pi}
\int_{C} \frac{dz\, f(z)}{\sinh^2(4 \pi g z)}
=
\int_{C} \frac{16 \pi g^2\,dz\, z^2}{\sinh^2(4 \pi g z)}\,
\lrbrk{1-\hypergeofn{2}{1}(-\half,\half;2;z^{-2} )}.
\]
The region encircled by the contour $C$
contains all the points $in/4g$ up to $n<4g/\epsilon$
and $f(z)$ must be holomorphic on it.
Let $\epsilon > 0$ infinitesimal. 
We shall consider integration over the closed contour,
see \figref{fig:contour}
\< \label{eq:contour}
C \eq C_{1} \cup C_{2} \cup C_{3} \cup C_{4} \cup C_{5}\, , 
\nln
C_{1}\eq (1/\epsilon)e^{i[0,\pi]}\, ,
\nln
C_{2} \eq 
[ - 1/\epsilon, -1-\epsilon ] \cup [+1+\epsilon,+1/\epsilon] \, ,
\nln
C_{3}\eq 
(-1+\epsilon e^{i[0,\pi]})\cup(+1+\epsilon e^{i[0,\pi]})\,,
\nln
C_{4} \eq 
[-1+\epsilon,-\epsilon] \cup [+\epsilon,+1-\epsilon] \, ,
\nln
C_{5}\eq \epsilon e^{i[0,\pi]}\,,
\>
Note that the hypergeometric function in $f(z)$ has 
singularities at $z=\pm 1$ and we shall deform the
standard cut on the interval $[-1,1]$
such that it lies below the contour $C_4$.

Let us first consider the integral over the large semicircle $C_{1}$. 
Here we have
\[
f(z)=\frac{1}{4}+\order{1/z}
\]
and for large enough $1/\epsilon$ we find
(neglecting corrections in $\epsilon$)
\[
8\pi g^2 \int_{C_{1}} 
\frac{dz\,f(z)}{\sinh^2(4 \pi g z)} 
=
2\pi g^2 \int_{C_{1}} 
\frac{dz}{\sinh^2(4 \pi g z)} 
=g \coth (4\pi g/\epsilon )
=g= c^{(0)}_{2,3} \,g.
\]

The first term in $f(z)$ leads to no singularity on the 
real axis and can be integrated 
straightforwardly over the remaining contours
\[
\lim_{m\to\infty}
\int_{C_2\cup C_3\cup C_4\cup C_5} \frac{16 \pi g^2\,dz\, z^2}{\sinh^2(4 \pi g z)}
=
\int_{-\infty}^\infty \frac{16 \pi g^2\,dz\, z^2}{\sinh^2(4 \pi g z)}
= \frac{1}{12 g}=\frac{c^{(2)}_{2,3}}{g}\,.
\]

For very small $z$ the remaining hypergeometric function 
goes as 
\[
\hypergeofn{2}{1}(-\half,\half;2;z^{-2} )\sim
 \frac{4 i}{3 \pi z}
\]
so that the semi-circle $C_5$ at the origin contributes 
\[
-\int_{C_5} \frac{16 \pi g^2\,dz\, z^2}{\sinh^2(4 \pi g z)}
\hypergeofn{2}{1}(-\half,\half;2;z^{-2} )=
-\frac{4}{3\pi}=
c_{2,3}^{(1)}\,.
\]

Since the integrand is an even function of $z$, the other three
pieces add up to the real part of an integral along the positive
semi-axis only:
\<
H(g) \eq -  16 \pi g^2 \int_{C_2\cup C_3\cup C_4}  \frac{ dz\,z^2 }{\sinh^2(4 \pi g z)} 
\, \hypergeofn{2}{1}(-\half,\half;2;z^{-2})
\nln\eq 
-  32 \pi g^2 \, \Re 
\int_0^\infty \frac{dz \, z^2}{\sinh^2(4 \pi g z)} 
\, \hypergeofn{2}{1}(-\half,\half;2;z^{-2})
\>
The real part of the hypergeometric function is easily read off
from its one-parameter representation
\[
\Re \hypergeofn{2}{1}(-\half,\half;2;z^{-2})
 = \frac{4}{\pi} \int_0^{\min(z,1)} dt\,
\sqrt{1-t^2}\sqrt{1 -t^2 z^{-2}  } \, .
\]
In $H(g)$ we readily estimate the contribution 
from $z>1$ to be of order $\order{e^{-g}}$
due to the suppression by $1/\sinh(4\pi g z)^2$.
The remainder is 
\[
H(g) = -  128 g^2 
\int_0^1 \frac{dz \, z^2}{\sinh^2(4 \pi g z)} 
 \int_0^{z} dt\,\sqrt{1-t^2}\sqrt{ 1 - t^2 z^{-2}  }
+\order{e^{-g}}
\,.
\]
To deal with these integrals we
successively
\begin{itemize}
\item expand $\sqrt{1-t^2}$ into a series in $t$
\[
\sqrt{1-t^2} = - \frac{1}{2 \sqrt{\pi}} 
\sum_{m=0}^\infty \frac{\gammafn(m-\frac{1}{2})}{\gammafn(m+1)}
\,t^{2n},
\]
\item calculate the $t$ integral in each term
\[
\int_0^{z} dt \,t^{2m} \sqrt{ 1 - t^2z^{-2}} 
= \frac{\sqrt{\pi}}{4} \, \frac{\gammafn(m + \half)}{\gammafn(m+2)} \,
z^{1 + 2 m},
\]
\item evaluate the $z$ integral. Here we may extend the range to the
whole positive semi-axis since the error is again seen to be
$\order{e^{-g}}$
\[
\int_0^{1} \frac{dz \,z^{2 m + 3}}{\sinh^2(4\pi g z)} =
\int_0^\infty \frac{dz \,z^{2 m + 3}}{\sinh^2(4\pi g z)}+\order{e^{-g}} = 
\frac{4\gammafn(2 m + 4) \, \zeta(2 m + 3)}{(8\pi g)^{2m+4}}+\order{e^{-g}}.
\]
\end{itemize}
On collecting terms:
\<
H(g) \eq \sum_{m=0}^\infty
\frac{\gammafn(m-\frac{1}{2}) \gammafn(m+\frac{1}{2})
\gammafn(2m+4)}{\gammafn(m+1) \gammafn(m+2)} \, \frac{\zeta(2 m +3)}{\pi^2(8 \pi g)^{2 m + 2}} 
+  \order{e^{-g}} 
\nln
 \eq \sum_{n=3}^\infty \frac{2\,\zeta(n)}{(-2\pi)^n \gammafn(n-1)} \, 
\frac{\gammafn[\frac{1}{2}(n+2)]\gammafn[\frac{1}{2}n]}{\gammafn[\frac{1}{2}(6-n)]\gammafn[\frac{1}{2}(4-n)]} 
\, g^{1-n}
+  \order{e^{-g}} 
\nln
\eq
\sum_{n=3}^\infty c_{2,3}^{(n)}\,g^{1-n}
+  \order{e^{-g}}
\, .
\>
The second line is obtained by using some identities of gamma functions
and setting $n=2m+3$.
In conclusion:
\[
c_{2,3}(g) =
c_{2,3}^{(0)}\,g+c_{2,3}^{(1)} +\frac{c^{(2)}_{2,3}}{g} + H(g)
= \sum_{m=0}^\infty c_{2,3}^{(m)} g^{1-m}  +  \order{e^{-g}}.
\]
We proved the correctness of the strong-coupling expansion
coefficients $c_{r,s}^{(m)}$ 
from equation \eqref{eq:coeffstrong} (see also \eqref{eq:c23zeta}) 
for the case $(r,s)=(2,3)$. The role of the zeta
function regularisation used in the main text is simply to dispose
of the terms of order $\order{e^{-g}}$ that become visible in this
proof.

\medskip

The argument presented in this appendix should remain applicable
for generic values of $(r,s)$.
We hope to return to this general case in future work.
We can however perform the analysis for the first two orders. 
By Euler-MacLaurin summation (or zeta function regularization), 
the leading two orders of $c_{r,s}(g)$ takes the form
\[
c^{(0)}_{r,s}=
\int_0^\infty h_{r,s}(1/z)\,dz,
\qquad
c^{(1)}_{r,s}=
\zeta(0)h_{r,s}(\infty).
\]
In fact, one can show that the above argument yields the same answers. 
The integral of $h_{r,s}$ in the form \eqref{hhyper} 
can be performed in closed form for all $(r,s)$
\[
c^{(0)}_{r,s}=
\frac{8(r-1)(s-1)\sin[\pi(r-s)]}{\pi (s-r-1)(s-r+1)(s+r-3)(s+r-1)}
=\delta_{r+1,s}
\]
and one obtains full agreement with the AFS phase 
\eqref{eq:coeff01} \cite{Arutyunov:2004vx}
using that $r,s$ are integers with $r\geq s+1$.
For the next order one can use the representation \eqref{cintrepsum}
of $h_{r,s}(z)$ at $z=\infty$
\<
c^{(1)}_{r,s}\eq 
\zeta(0)
2\cos[\half\pi(s-r-1)]\,(r-1)(s-1)
\int_0^\infty \frac{dt}{t}\,\besselJ{r-1}(2\,t)\besselJ{s-1}(2\,t)
\nln\eq
-\cos[\half\pi(s-r-1)](r-1)(s-1)
\frac{2\cos[\half\pi(s-r-1)]}{\pi (s-r)(s+r-2)}
\nln\eq
-\frac{1-(-1)^{r+s}}{\pi}\,\frac{(r-1)(s-1)}{(s-r)(s+r-2)}\,.
\>
This is again in full agreement with the HL phase 
\eqref{eq:coeff01} \cite{Hernandez:2006tk}.

\bibliographystyle{nb}
\bibliography{gaugephase}

\end{document}